\documentclass[onecolumn, 12pt, draft]{IEEEtran}

\usepackage{amsmath}
\usepackage{amsthm}
\usepackage{amssymb}
\usepackage{cite}
\usepackage{amsfonts}
\usepackage{enumerate}
\usepackage{subfigure}
\usepackage{multirow}
\usepackage{verbatim}
\usepackage[final]{lscape,graphicx}

\title{Unified Sum-BER Performance Analysis of AF MIMO Beamforming in Two-Way Relay Networks}
\author{Hyunjun Kim and Cihan Tepedelenlio\u{g}lu\\
\small{School of Electrical, Computer, and Energy Engineering,
Arizona State University, Tempe, AZ 85287-5706}\\
\small{\{Hyunjun.Kim, Cihan\}@asu.edu}}

\begin{document}

\maketitle

\begin{abstract}

Unified performance analysis is carried out for amplify-and-forward (AF) multiple-input-multiple-output (MIMO) beamforming (BF) two-way relay networks in Rayleigh fading with five different relaying protocols including two novel protocols for better performance. As a result, a novel closed-form sum-bit error rate (BER) expression is presented in a unified expression for all protocols. A new closed-form high signal-to-noise-ratio (SNR) performance is also obtained in a single expression, and an analytical high-SNR gap expression between the five protocols is provided. We compare the performance of the five relaying protocols with respect to sum-BER with appropriately normalized rate and power, and show that the proposed protocol with four time slots outperforms other protocols when transmit powers from two sources are sufficiently different, and the one with three time slots dominates other protocols when multiple relay antennas are used, at high-SNR.

\end{abstract}


\section{Introduction}

Multiple-input multiple-output (MIMO) technology has been considered as a way to combat severe fading due to its excellent link reliability based on achievable spatial diversity \cite{paulraj04}. When multiple antennas are used, the combination of maximum ratio transmission (MRT) beamforming (BF) \cite{lo99}, and maximum ratio combining (MRC) beamforming \cite{goldsmith05} is one simple way to achieve spatial diversity. Cooperative diversity schemes, using relays between the source and destination, have been widely investigated because of their spatial diversity and extensive coverage with reduced power consumption \cite{laneman04, send103}. Amplify-and-forward (AF) relaying using two time slots is known to offer gains in performance, in which the source transmits its signal to the relay in the first time slot, and the relay amplifies and forwards the transmitted signal to the destination in the second time slot \cite{laneman04, send103}. We refer to this scheme as ``one-way relaying" to distinguish from ``two-way relaying" which is our focus.

Even though one-way relaying provides spatial diversity and extensive coverage with reduced power consumption, it causes a spectral loss due to the use of more time slots when two sources \(\rm A\) and \(\rm B\) communicate each other through a relay \(\rm R\), as in Figure 1. To improve the spectral efficiency using two time slots, two-way relaying is suggested, in which two sources transmit simultaneously their signals to the relay in the first time slot (multiple access phase), and the relay amplifies received signals and forwards the combined signals to the sources in the second time slot (broadcast phase) \cite{rankov07, popovski07, louie10}.

After AF and decode-and-forward (DF) two-way relay networks are proposed in \cite{popovski07}, sum-bit error rate (BER) and maximum ergodic sum-rate for systems using a single antenna at all nodes are analyzed for two-way relay systems in \cite{han09, louie10, guo11}. Reference \cite{louie10} provides closed-form sum-BER and maximum ergodic sum-rate for the two-slot, three-slot, and four-slot two-way relay systems with a single antenna at each node over Rayleigh fading, and introduces power allocation at \(\rm R\) for the three-slot protocol, which proves useful when average transmit SNRs at \(\rm A\) and \(\rm B\) are sufficiently different (i.e. ``unbalanced"). Reference \cite{han09} also presents sum-BER and maximum ergodic sum-rate bounds for systems using Alamouti code for the two-slot protocol when multiple antennas are used at \(\rm A\) and \(\rm B\) while a single antenna is used at \(\rm R\). Performance analysis is carried out for AF two-slot two-way relay systems with BF using a single relay antenna over Nakagami-\textit{m} fading in \cite{guo11_1}. Using multiple antennas at \(\rm R\), meanwhile, BF optimization for only maximum ergodic sum-rate is conducted without performance analysis for AF MIMO two-slot two-way relay systems in \cite{lee08, lee10, wang10}. {\it BF optimization} is our term for simultaneous beamforming at \(\rm R\) to both \(\rm A\) and \(\rm B\). Reference \cite{panah10} investigates the effects of channel estimation error at \(\rm A\) and \(\rm B\) for AF MIMO two-way relaying, and provides maximum ergodic sum-rate lower-bounds with imperfect channel state information (CSI) at \(\rm A\) and \(\rm B\).

Based on this background, our contributions are as follows:
\begin{itemize}
\item Novel closed-form sum-BER expressions are presented in a unified framework for five AF MIMO two-way relaying protocols with BF.
\item This is the first paper dealing with performance analysis of AF MIMO two-way relay networks using BF with multiple relay antennas, to the best of our knowledge.
\item Two novel two-way relaying protocols are proposed using three or four time slots, and we show that two proposed protocols outperform existing protocols in sum-BER at high-SNR.
\item New closed-form high-SNR sum-BER expressions are provided in a single formula for all five AF MIMO BF two-way relaying protocols. Based on this high-SNR analysis, an analytical high-SNR gap expression between the five different protocols is provided, taking into account the appropriate rate and power normalization.
\end{itemize}

After system models are described for the five two-way relaying protocols with a single relay antenna in Section II, unified performance analysis including high-SNR analysis is presented in Section III. Multiple relay antennas are considered in Section IV. Numerical and Monte-Carlo simulations compare the performance of five different relaying protocols in Section V. Conclusions are given in Section VI.

\section{System Model}

Figure 1 shows a two-hop MIMO two-way relay system, which consists of two sources, which are also destinations, \(\rm A\) and \(\rm B\), and a relay \(\rm R\). All nodes are equipped with multiple antennas, \(M_{\rm A}\), \(M_{\rm B}\), and \(M_{\rm R}\), respectively. \(\mathbf{H}_{\rm AR}\) and \(\mathbf{H}_{\rm BR}\) are \(M_{\rm R}\times M_{\rm A}\) and \(M_{\rm R}\times M_{\rm B}\) statistically independent complex Gaussian channel matrices connecting the nodes, respectively. The channel coefficients are assumed to remain static while \(\rm A\) and \(\rm B\) exchange their data, and channels are reciprocal in the sense that \(\mathbf{H}_{\rm RA} = \mathbf{H}_{\rm AR}^H\) (\(M_{\rm A}\times M_{\rm R}\)) and \(\mathbf{H}_{\rm RB} = \mathbf{H}_{\rm BR}^H\) (\(M_{\rm B}\times M_{\rm R}\)), where \((\cdot)^H\) denotes a matrix Hermitian. We assume that transmitters have CSI only on connected nodes while receivers can access full CSI.

A half-duplex time division multiple access (TDMA) scenario is considered with five different transmission protocols, illustrated in Figure 2. In this work, the direct links, \(\rm A \rightarrow B\) and \(\rm B \rightarrow A\), are assumed to be negligible even though their presence can be incorporated into our analysis. Symbols are transmitted with zero mean and unit variance, and additive noise is independent complex Gaussian with zero mean and unit variance. When multiple antennas are considered at \(\rm R\), BF optimization has to be conducted at \(\rm R\) in the two-slot and first three-slot protocols, where \(\rm R\) beamforms to \(\rm A\) and \(\rm B\) simultaneously. We therefore first consider a single relay antenna to obtain closed-form expressions for all protocols in Section III, and extend this to multiple antennas in Section IV. In what follows, we present unified instantaneous received SNR representations for each protocol. Note that when the protocols with different number of slots are compared, transmit power is normalized so that each node uses the same power, and the constellation sizes are chosen so that the rates are fixed as well.

\subsection{Extension of Existing Protocols}

In this subsection, three two-way relaying protocols discussed in \cite{louie10}, where only a single antenna is considered at all nodes, are extended to multiple antennas with BF at \(\rm A\) and \(\rm B\). Note that BF optimization is not necessary even for the two-slot and first three-slot protocols when \(M_{\rm R} = 1\), so that performance analysis in closed-form is tractable.

\subsubsection{Two-Slot Protocol}

In the two-slot protocol, \(\rm A\) and \(\rm B\) transmit their signals to \(\rm R\) using the corresponding matched BF vectors in the first time slot, and \(\rm R\) amplifies the sum as in \cite{han09, louie10, guo11} and forwards it to \(\rm A\) and \(\rm B\) in the second time slot. When \(\rm A\) and \(\rm B\) beamform in the first time slot, they use the so-called {\it matched} BF vectors, the strongest right singular vectors of \(\mathbf{H}_{\rm AR}\) and \(\mathbf{H}_{\rm BR}\), denoted by \(\mathbf{f}_{\rm AR}\) and \(\mathbf{f}_{\rm BR}\), respectively.

\subsubsection{First Three-Slot Protocol}

In the first three-slot protocol, \(\rm A\) transmits its signal to \(\rm R\) using \(\mathbf{f}_{\rm AR}\) in the first time slot; \(\rm B\) transmits its signal to \(\rm R\) using \(\mathbf{f}_{\rm BR}\) in the second time slot; \(\rm R\) weighs the received signals from \(\rm A\) and \(\rm B\) with coefficients \(\alpha \ge 0\) and \(\beta \ge 0\) satisfying \(\alpha^2 + \beta^2 = 1\), amplifies the weighted sum, and forwards it to \(\rm A\) and \(\rm B\) in the third time slot. Coefficients \(\alpha\) and \(\beta\) are weights for two received signals from \(\rm A\) and \(\rm B\) at \(\rm R\), respectively, which can be determined to minimize instantaneous sum-BERs using brute force search \cite{louie10}. Since there is no closed-form for \(\alpha\) and \(\beta\) when instantaneous sum-BER is optimized, \(\alpha\) and \(\beta\) can also be chosen based on average channel statistics using our high-SNR expressions, as described in Section III.C.

\subsubsection{First Four-Slot Protocol (One-Way Relaying)}

In the first four-slot protocol, \(\rm A\) transmits its signal to \(\rm R\) using \(\mathbf{f}_{\rm AR}\) in the first time slot; \(\rm R\) amplifies the received signal and forwards it to \(\rm B\) in the second time slot; \(\rm B\) transmits its signal using \(\mathbf{f}_{\rm AR}\) to \(\rm R\) in the third time slot; \(\rm R\) amplifies the other received signal and forwards it to \(\rm A\) in the fourth time slot. This amounts to one-way relaying sequentially, \(\rm A\rightarrow R\rightarrow B\) and \(\rm B\rightarrow R\rightarrow A\). Note that power normalization is required due to two transmissions at \(\rm R\) (i.e. half of the power used by the two-slot protocol).

\subsection{Proposed Protocols}

In what follows, we propose new relaying protocols for better performance in closed-form.

\subsubsection{Second Three-Slot Protocol}

In the second three-slot protocol, \(\rm A\) and \(\rm B\) transmit their signals using \(\mathbf{f}_{\rm AR}\) and \(\mathbf{f}_{\rm BR}\), respectively, to \(\rm R\) in the first time slot, \(\rm R\) amplifies the received sum and forwards it to \(\rm A\) and \(\rm B\) in the second and third time slots, consecutively, and both signals are received at \(\rm A\) and \(\rm B\). Since \(\rm R\) forwards twice, transmit power normalization is required at \(\rm R\). To combine the two received signals at the receivers, the minimum mean square error (MMSE) combining scheme is used \cite{kim09_a, kim10_i, kim10_a}. Note that there is no need for combining at the destination (i.e. \(\rm A\) or \(\rm B\)) in the existing protocols of Section II.A since the desired signals are only received once at the destination.

\subsubsection{Second Four-Slot Protocol}

The second four-slot protocol is proposed to obtain better sum-BER by taking advantage of the technique used in the first three-slot protocol, which is weighting two received signals from \(\rm A\) and \(\rm B\) at \(\rm R\) with \(\alpha\) and \(\beta\), respectively. In the second four-slot protocol, \(\rm A\) transmits its signal using \(\mathbf{f}_{\rm AR}\) to \(\rm R\) in the first time slot; \(\rm B\) transmits its signal using \(\mathbf{f}_{\rm BR}\) to \(\rm R\) in the second time slot; \(\rm R\) weighs the received signals with coefficients \(\alpha\) and \(\beta\), amplifies the weighted sum and forwards it to \(\rm A\) and \(\rm B\) in the third and fourth time slots, consecutively. Transmit power normalization is also required at \(\rm R\) due to two transmissions. To combine two received signals at \(\rm A\) and \(\rm B\), separately, MMSE combining is used \cite{kim09_a, kim10_i, kim10_a}.

\subsection{Unified SNR Representations for Five Different Protocols for $M_{\rm R} = 1$}

For the aforementioned protocols, after canceling the self-interferences as in \cite{han09, louie10, guo11}, portions of received signals coming back through \(\rm R\) induced by \(\rm A\) and \(\rm B\), with MRC and MMSE combining, the instantaneous received SNRs at \(\rm A\) and \(\rm B\) can be expressed, respectively, in a unified framework:
\begin{footnotesize}\begin{equation}
\gamma_{\rm BRA} = \frac{A_{\rm BRA}\gamma_{\rm BR}\gamma_{\rm RA}}{B_{\rm BRA}\gamma_{\rm BR}+ C_{\rm BRA}\gamma_{\rm RA} + 1}
\end{equation}
\begin{equation}
\gamma_{\rm ARB} = \frac{A_{\rm ARB}\gamma_{\rm AR}\gamma_{\rm RB}}{B_{\rm ARB}\gamma_{\rm AR}+ C_{\rm ARB}\gamma_{\rm RB} + 1},
\end{equation}\end{footnotesize}\noindent
where \(\gamma_{\rm AR} = \rho_{\rm AR}\Vert{\mathbf{h}_{\rm AR}\mathbf{f}_{\rm AR}}\Vert^2\), \(\gamma_{\rm BR} = \rho_{\rm BR}\Vert{\mathbf{h}_{\rm BR}\mathbf{f}_{\rm BR}}\Vert^2\), \(\gamma_{\rm RA} = \rho_{\rm RA}\Vert{\mathbf{h}_{\rm RA}}\Vert^2,\) and \(\gamma_{\rm RB} = \rho_{\rm RB}\Vert{\mathbf{h}_{\rm RB}}\Vert^2\); \(\rho_{\rm AR},\) \(\rho_{\rm BR},\) \(\rho_{\rm RA}\), and \(\rho_{\rm RB}\) are average transmit SNRs, where we assume \(\rho_{\rm RA} = \rho_{\rm RB}\); \(\mathbf{h}_{\rm AR}\), \(\mathbf{h}_{\rm BR}\), \(\mathbf{h}_{\rm RA} = \mathbf{h}_{\rm AR}^H\), and \(\mathbf{h}_{\rm RB} = \mathbf{h}_{\rm BR}^H\) are channel coefficient vectors, assumed to be i.i.d. \(\cal{CN}\)\((0, 1)\); \(\mathbf{f}_{\rm AR}\) and \(\mathbf{f}_{\rm BR}\) are BF vectors with norm \(1\) obtained as \(\mathbf{h}_{\rm AR}^H/\Vert{\mathbf{h}_{\rm AR}}\Vert\) and \(\mathbf{h}_{\rm BR}^H/\Vert{\mathbf{h}_{\rm BR}}\Vert\), respectively; \(A_{\rm BRA}\), \(B_{\rm BRA}\), \(C_{\rm BRA}\), \(A_{\rm ARB}\), \(B_{\rm ARB}\), and \(C_{\rm ARB}\) are non-negative constants given in Table I for all five protocols. These SNR representations will be used to find distributions for performance analysis. We consider removing 1 from equations (1) and (2) to obtain closed-form sum-BER expressions, denoted respectively as \(\Gamma_{\rm BRA}\) and \(\Gamma_{\rm ARB}\), which are equivalent to equations (1) and (2) at high-SNR \cite{louie10}.

\section{Performance Analysis for $M_{\rm R} = 1$}

Sum-BER performance analysis including high-SNR analysis is carried out using the unified received SNR expressions. The multiple relay antenna case, $M_{\rm R} > 1$, is described in Section IV.

\subsection{Performance Metric}

For the performance metric, we consider sum-BER, sum of BERs at \(\rm A\) and \(\rm B\), since there are two receiving nodes and the worse one dominates the sum and closely approximates the worst of the two BERs. Sum-BER for all protocols is defined as follows:
\begin{footnotesize}\begin{equation}
P_{\rm b} = \frac{1}{\log_2(M)}\int_0^{\infty}aQ\left(\sqrt{2bx}\right)\left(f_{\gamma_{\rm ARB}}(x) + f_{\gamma_{\rm BRA}}(x)\right)dx,
\end{equation}\end{footnotesize}\noindent
where \(Q(x) := \left(1/\sqrt{2\pi}\right)\int_x^{\infty}e^{-y^2/2}dy\) and \(a\) and \(b\) are modulation related positive constants. For example, \(a = 1\) and \(b = 1\) provide exact BER for binary phase shift keying (BPSK), while \(a = 2\) and \(b = \sin^2(\pi/M)\) and \(a = 4\left(1-1/\sqrt{M}\right)\) and \(b = 3/(2(M-1))\) provide tight SER approximations for $M$-ary PSK ($M$-PSK) and $M$-ary quadrature amplitude modulation ($M$-QAM), respectively.

\subsection{Sum-BER using Unified SNR Representations}

When cumulative distribution functions (CDFs) are available instead of probability density functions (PDFs), the following alternative equation can be used to calculate sum-BER.
\begin{footnotesize}\begin{equation}\begin{split}
&P_{\rm b} = \frac{a\sqrt{b}}{2\sqrt{\pi}\log_2(M)}\int_0^{\infty}\frac{e^{-bx}}{\sqrt{x}}\left(F_{\gamma_{\rm BRA}}(x) + F_{\gamma_{\rm ARB}}(x)\right)dx\\
&\hspace{0.15 in}\ge \frac{a\sqrt{b}}{2\sqrt{\pi}\log_2(M)}\int_0^{\infty}\frac{e^{-bx}}{\sqrt{x}}\left(F_{\Gamma_{\rm BRA}}(x) + F_{\Gamma_{\rm ARB}}(x)\right)dx.
\end{split}\end{equation}\end{footnotesize}\noindent
Note that the second line of equation (4) provides a lower-bound in sum-BER since the CDFs of \(\Gamma_{\rm BRA}\) and \(\Gamma_{\rm ARB}\), described at the end of Section II.C, are used.

To calculate sum-BER using the unified SNR representations, the distributions of equations (1) and (2) should be obtained first. Since we use the distributions of \(\Gamma_{\rm BRA}\) and \(\Gamma_{\rm ARB}\), when we consider Rayleigh fading, the distributions can be obtained as follows (please see Appendix I for derivations):
\begin{footnotesize}\begin{equation}\begin{split}
&F_{\Gamma_{\rm BRA}}(x) = 1 - \sum_{p=0}^{M_{\rm B}-1}\sum_{k=0}^{M_{\rm A}+p-1}\dbinom{M_{\rm A}+p-1}{k}
\frac{2B_{\rm BRA}^{\frac{2M_{\rm A}+p-k-1}{2}}C_{\rm BRA}^{\frac{k+p+1}{2}}}{A_{\rm BRA}^{M_{\rm A+p}}p!\left(M_{\rm A}-1\right)!\rho_{\rm BR}^{\frac{k+p+1}{2}}\rho_{\rm RA}^{\frac{2M_{\rm A}+p-k-1}{2}}}\\
&\hspace{0.75 in}x^{M_{\rm A}+p}e^{-\frac{x}{A_{\rm BRA}}\left(\frac{C_{\rm BRA}}{\rho_{\rm BR}}+\frac{B_{\rm BRA}}{\rho_{\rm RA}}\right)}K_{k-p+1}\left(\frac{2x}{A_{\rm BRA}}\sqrt{\frac{B_{\rm BRA}C_{\rm BRA}}{\rho_{\rm BR}\rho_{\rm RA}}}\right),
\end{split}\end{equation}
\begin{equation}\begin{split}
&F_{\Gamma_{\rm ARB}}(x) = 1 - \sum_{p=0}^{M_{\rm A}-1}\sum_{k=0}^{M_{\rm B}+p-1}\dbinom{M_{\rm B}+p-1}{k}
\frac{2B_{\rm ARB}^{\frac{2M_{\rm B}+p-k-1}{2}}C_{\rm ARB}^{\frac{k+p+1}{2}}}{A_{\rm ARB}^{M_{\rm B+p}}p!\left(M_{\rm B}-1\right)!\rho_{\rm AR}^{\frac{k+p+1}{2}}\rho_{\rm RB}^{\frac{2M_{\rm B}+p-k-1}{2}}}\\
&\hspace{0.75 in}x^{M_{\rm B}+p}e^{-\frac{x}{A_{\rm ARB}}\left(\frac{C_{\rm ARB}}{\rho_{\rm AR}}+\frac{B_{\rm ARB}}{\rho_{\rm RB}}\right)}K_{k-p+1}\left(\frac{2x}{A_{\rm ARB}}\sqrt{\frac{B_{\rm ARB}C_{\rm ARB}}{\rho_{\rm AR}\rho_{\rm RB}}}\right),
\end{split}\end{equation}\end{footnotesize}\noindent
where \(K_{\nu}(x)\) is the modified Bessel function of the second kind \cite{gradshteyn07}.

Since the CDFs of \(\Gamma_{\rm BRA}\) and \(\Gamma_{\rm ARB}\) are mathematically tractable, the alternative expression in equation (4) can be used to calculate sum-BER. As a result, once equations (5) and (6) are substituted to the second line of equation (4), the sum-BER can be lower-bounded in closed-form as
\begin{footnotesize}\begin{equation}\begin{split}
&P_{\rm b} \ge \frac{a}{\log_2(M)} - \sum_{p=0}^{M_{\rm A}-1}\sum_{k=0}^{M_{\rm B}+p-1}\dbinom{M_{\rm B}+p-1}{k}
\frac{a\sqrt{b}B_{\rm ARB}^{\frac{2M_{\rm B}+p-k-1}{2}}C_{\rm ARB}^{\frac{k+p+1}{2}}}{\log_2(M)A_{\rm ARB}^{M_{\rm B+p}}p!\left(M_{\rm B}-1\right)!\rho_{\rm AR}^{\frac{k+p+1}{2}}\rho_{\rm RB}^{\frac{2M_{\rm B}+p-k-1}{2}}}\\
&\hspace{0.27 in}\frac{\left(\frac{4}{A_{\rm ARB}}\sqrt{\frac{B_{\rm ARB}C_{\rm ARB}}{\rho_{\rm AR}\rho_{\rm RB}}}\right)^{k-p+1}}{\left(b+\frac{C_{\rm ARB}}{A_{\rm ARB}\rho_{\rm AR}}+\frac{B_{\rm ARB}}{A_{\rm ARB}\rho_{\rm RB}}+\frac{2}{A_{\rm ARB}}\sqrt{\frac{B_{\rm ARB}C_{\rm ARB}}{\rho_{\rm AR}\rho_{\rm RB}}}\right)^{M_{\rm B}+k+\frac{3}{2}}}\frac{\Gamma\left(M_{\rm B}+k+\frac{3}{2}\right)\Gamma\left(M_{\rm B}+2p-k-\frac{1}{2}\right)}{\Gamma(M_{\rm B}+p+1)} \\
&\hspace{0.27 in}_2F_1\left(M_{\rm B}+k+\frac{3}{2},k-p+\frac{3}{2};M_{\rm B}+p+1;\frac{b+\frac{C_{\rm ARB}}{A_{\rm ARB}\rho_{\rm AR}}+\frac{B_{\rm ARB}}{A_{\rm ARB}\rho_{\rm RB}}-\frac{2}{A_{\rm ARB}}\sqrt{\frac{B_{\rm ARB}C_{\rm ARB}}{\rho_{\rm AR}\rho_{\rm RB}}}}{b+\frac{C_{\rm ARB}}{A_{\rm ARB}\rho_{\rm AR}}+\frac{B_{\rm ARB}}{A_{\rm ARB}\rho_{\rm RB}}+\frac{2}{A_{\rm ARB}}\sqrt{\frac{B_{\rm ARB}C_{\rm ARB}}{\rho_{\rm AR}\rho_{\rm RB}}}}\right) \\
&\hspace{0.27 in}- \sum_{p=0}^{M_{\rm B}-1}\sum_{k=0}^{M_{\rm A}+p-1}\dbinom{M_{\rm A}+p-1}{k}
\frac{a\sqrt{b}B_{\rm BRA}^{\frac{2M_{\rm A}+p-k-1}{2}}C_{\rm BRA}^{\frac{k+p+1}{2}}}{\log_2(M)A_{\rm BRA}^{M_{\rm A+p}}p!\left(M_{\rm A}-1\right)!\rho_{\rm BR}^{\frac{k+p+1}{2}}\rho_{\rm RA}^{\frac{2M_{\rm A}+p-k-1}{2}}}\\
&\hspace{0.25 in}\frac{\left(\frac{4}{A_{\rm BRA}}\sqrt{\frac{B_{\rm BRA}C_{\rm BRA}}{\rho_{\rm BR}\rho_{\rm RA}}}\right)^{k-p+1}}{\left(b+\frac{C_{\rm BRA}}{A_{\rm BRA}\rho_{\rm BR}}+\frac{B_{\rm BRA}}{A_{\rm BRA}\rho_{\rm RA}}+\frac{2}{A_{\rm BRA}}\sqrt{\frac{B_{\rm BRA}C_{\rm BRA}}{\rho_{\rm BR}\rho_{\rm RA}}}\right)^{M_{\rm A}+k+\frac{3}{2}}}\frac{\Gamma\left(M_{\rm A}+k+\frac{3}{2}\right)\Gamma\left(M_{\rm A}+2p-k-\frac{1}{2}\right)}{\Gamma(M_{\rm A}+p+1)} \\
&\hspace{0.27 in}_2F_1\left(M_{\rm A}+k+\frac{3}{2},k-p+\frac{3}{2};M_{\rm A}+p+1;\frac{b+\frac{C_{\rm BRA}}{A_{\rm BRA}\rho_{\rm BR}}+\frac{B_{\rm BRA}}{A_{\rm BRA}\rho_{\rm RA}}-\frac{2}{A_{\rm BRA}}\sqrt{\frac{B_{\rm BRA}C_{\rm BRA}}{\rho_{\rm BR}\rho_{\rm RA}}}}{b+\frac{C_{\rm BRA}}{A_{\rm BRA}\rho_{\rm BR}}+\frac{B_{\rm BRA}}{A_{\rm BRA}\rho_{\rm RA}}+\frac{2}{A_{\rm BRA}}\sqrt{\frac{B_{\rm BRA}C_{\rm BRA}}{\rho_{\rm BR}\rho_{\rm RA}}}}\right),
\end{split}\end{equation}\end{footnotesize}\noindent
where \(_2F_1(\alpha,\beta;\gamma;z)\) is the Gauss hypergeometric function \cite[p.1005]{gradshteyn07}. To obtain equation (7), the following integral is used \cite[p.700]{gradshteyn07}:
\begin{footnotesize}\begin{equation}
\int_0^\infty x^{\mu-1}e^{-\alpha x}K_{\nu}(\beta x)dx = \frac{\sqrt{\pi}\left(2\beta\right)^{\nu}}{\left(\alpha+\beta\right)^{\mu+\nu}}\frac{\Gamma(\mu+\nu)\Gamma(\mu-\nu)}{\Gamma(\mu+\frac{1}{2})}
\hspace{0 in}_2F_1\left(\mu+\nu,\nu+\frac{1}{2};\mu+\frac{1}{2};\frac{\alpha-\beta}{\alpha+\beta}\right).
\end{equation}\end{footnotesize}\noindent
Note that equation (7) provides tight sum-BER lower-bounds for all five two-way relay protocols.

\subsection{High-SNR Analysis for Sum-BER using Unified SNR Representations}

The expression in equation (7) is tight at high SNR, but rather complicated. Simple high-SNR performance is now considered to simplify considerably by diversity and array gain analysis. The approximation uses the probability density functions (PDFs) of instantaneous SNRs normalized by the average SNR on each link defined as \(\lambda_{\rm ARB} := \Gamma_{\rm ARB}/\rho_{\rm AR}\) and \(\lambda_{\rm BRA} := \Gamma_{\rm BRA}/\rho_{\rm AR}\). The PDFs of \(\lambda_{\rm ARB}\) and \(\lambda_{\rm BRA}\) are shown satisfying the assumptions in \cite{wang03}, which provides a systematic method for high-SNR analysis. To simplify our analysis, we assume that \(\rho_{\rm BR}\), \(\rho_{\rm RA}\), and \(\rho_{\rm RB}\) are constant multiples of \(\rho_{\rm AR}\). Based on \cite[eqn.(1)]{wang03}, the average sum-BER of an uncoded system can be written as
\begin{footnotesize}\begin{equation}
P_{\rm b} = \frac{1}{\log_2(M)}\left(\left(2b\rho_{\rm AR}G_{{\rm ARB}}\right)^{-d_{{\rm ARB}}} + \left(2b\rho_{\rm AR}G_{{\rm BRA}}\right)^{-d_{{\rm BRA}}}\right) + o\left(\rho_{\rm AR}^{-\min\{d_{{\rm ARB}}, d_{{\rm BRA}}\}}\right),
\end{equation}\end{footnotesize}\noindent
as \(\rho_{\rm AR} \rightarrow \infty\), where \(G_{{\rm ARB}} = \left(\sqrt{\pi}\left(t_{\rm ARB}+1\right)/\left(a2^{t_{\rm ARB}}\eta_{\rm ARB}\Gamma\left(t_{\rm ARB}+3/2\right)\right)\right)^{1/(t_{\rm ARB}+1)}\) and \(G_{{\rm BRA}} = \left(\sqrt{\pi}\left(t_{\rm BRA}+1\right)/\left(a2^{t_{\rm BRA}}\eta_{\rm BRA}\Gamma\left(t_{\rm BRA}+3/2\right)\right)\right)^{1/(t_{\rm BRA}+1)}\) are the array gains; \(d_{{\rm ARB}} = t_{\rm ARB} + 1\) and \(d_{{\rm BRA}} = t_{\rm BRA} + 1\) are the diversity orders; \(t_{\rm ARB}\) and \(t_{\rm BRA}\) are the first nonzero derivative orders of the PDFs of channel dependent random variables, \(\lambda_{\rm ARB}\) and \(\lambda_{\rm BRA}\), at the origin, respectively; \(\eta_{\rm ARB} = f_{\lambda_{\rm ARB}}^{\left(t_{\rm ARB}\right)}(0)/\Gamma\left(t_{\rm ARB}+1\right)\not= 0\) and \(\eta_{\rm BRA} = f_{\lambda_{\rm BRA}}^{\left(t_{\rm BRA}\right)}(0)/\Gamma\left(t_{\rm BRA}+1\right)\not= 0\). Therefore, equation (9) can be calculated once \(t_{\rm ARB}\), \(t_{\rm BRA}\), \(\eta_{\rm ARB}\), and \(\eta_{\rm BRA}\) are found.

For the \(\rm A \rightarrow R \rightarrow B\) path, \(t_{\rm ARB} = \min\{M_{\rm A}, M_{\rm B}\} - 1\) since the diversity order of the \(\rm A \rightarrow R \rightarrow B\) path is \(\min\{M_{\rm A}, M_{\rm B}\}\) \cite[eqn.(16)]{kim10_a}. The \(t_{\rm ARB}\) order derivative of the PDF of \(\lambda_{\rm ARB}\) evaluated at the origin can be obtained as (please see Appendix II for derivation)
\begin{footnotesize}\begin{equation}
f_{\lambda_{\rm ARB}}^{(t_{\rm ARB})}(0) =
\begin{cases} f_{\lambda_{\rm RB}}^{(t_{\rm RB})}(0), \hspace{0.8 in} M_{\rm A}>M_{\rm B}\\
f_{\lambda_{\rm AR}}^{(t_{\rm AR})}(0), \hspace{0.8 in} M_{\rm A}<M_{\rm B}\\
f_{\lambda_{\rm AR}}^{(t_{\rm AR})}(0) + f_{\lambda_{\rm RB}}^{(t_{\rm RB})}(0), \hspace{.15 in} M_{\rm A}=M_{\rm B}
\end{cases},
\end{equation}\end{footnotesize}\noindent
where \(t_{\rm AR} = M_{\rm A} - 1\) and \(t_{\rm RB} = M_{\rm B} - 1\) \cite[eqn.(12)]{kim10_a} are the first nonzero derivative orders of the PDFs of \(\lambda_{\rm AR} := \gamma_{\rm AR}/\rho_{\rm AR}\) and \(\lambda_{\rm RB} := \gamma_{\rm RB}/\rho_{\rm AR}\), at the origin, respectively;
\begin{footnotesize}\begin{equation}
f_{\lambda_{\rm AR}}^{(t_{\rm AR})}(0) = \left(\frac{C_{\rm ARB}}{A_{\rm ARB}}\right)^{t_{\rm AR}+1} \hspace{0.25 in}
\end{equation}
\begin{equation}
f_{\lambda_{\rm RB}}^{(t_{\rm RB})}(0) = \left(\frac{B_{\rm ARB}\rho_{\rm AR}}{A_{\rm ARB}\rho_{\rm RB}}\right)^{t_{\rm RB}+1}.
\end{equation}\end{footnotesize}\noindent
Therefore, \(\eta_{\rm ARB}\) can be written as
\begin{footnotesize}\begin{equation}
\eta_{\rm ARB} = \frac{f_{\lambda_{\rm ARB}}^{(t_{\rm ARB})}(0)}{\Gamma\left(\min\{M_{\rm A}, M_{\rm B}\}\right)}.
\end{equation}\end{footnotesize}\noindent

Similarly, for the \(\rm B \rightarrow R \rightarrow A\) path, \(t_{\rm BRA} = \min\{M_{\rm A}, M_{\rm B}\} - 1\) since the diversity order of the \(\rm B \rightarrow R \rightarrow A\) path is \(\min\{M_{\rm A}, M_{\rm B}\}\). The \(t_{\rm BRA}\) order derivative of the PDF of \(\lambda_{\rm BRA}\) evaluated at the origin can be obtained as (please see Appendix II for derivation)
\begin{footnotesize}\begin{equation}
f_{\lambda_{\rm BRA}}^{(t_{\rm BRA})}(0) =
\begin{cases} f_{\lambda_{\rm BR}}^{(t_{\rm BR})}(0), \hspace{0.8 in} M_{\rm A}>M_{\rm B}\\
f_{\lambda_{\rm RA}}^{(t_{\rm RA})}(0), \hspace{0.8 in} M_{\rm A}<M_{\rm B}\\
f_{\lambda_{\rm BR}}^{(t_{\rm BR})}(0) + f_{\lambda_{\rm RA}}^{(t_{\rm RA})}(0), \hspace{.15 in} M_{\rm A}=M_{\rm B}
\end{cases},
\end{equation}\end{footnotesize}\noindent
where \(t_{\rm BR} = M_{\rm B} - 1\), \(t_{\rm RA} = M_{\rm A} - 1\) \cite[eqn.(12)]{kim10_a} are the first nonzero derivative orders of the PDFs of \(\lambda_{\rm BR} := \gamma_{\rm BR}/\rho_{\rm AR}\) and \(\lambda_{\rm RA} := \gamma_{\rm RA}/\rho_{\rm AR}\), at the origin, respectively;
\begin{footnotesize}\begin{equation}
f_{\lambda_{\rm BR}}^{(t_{\rm BR})}(0) = \left(\frac{C_{\rm BRA}\rho_{\rm AR}}{A_{\rm BRA}\rho_{\rm BR}}\right)^{t_{\rm BR}+1} \hspace{0.05 in}
\end{equation}
\begin{equation}
f_{\lambda_{\rm RA}}^{(t_{\rm RA})}(0) = \left(\frac{B_{\rm BRA}\rho_{\rm AR}}{A_{\rm BRA}\rho_{\rm RA}}\right)^{t_{\rm RA}+1}.
\end{equation}\end{footnotesize}\noindent
Therefore, \(\eta_{\rm BRA}\) can be written as
\begin{footnotesize}\begin{equation}
\eta_{\rm BRA} = \frac{f_{\lambda_{\rm BRA}}^{(t_{\rm BRA})}(0)}{\Gamma\left(\min\{M_{\rm A}, M_{\rm B}\}\right)}.
\end{equation}\end{footnotesize}\noindent

As a consequence, high-SNR performance can be obtained as follows:
\begin{footnotesize}\begin{equation}
P_{\rm b} = \frac{1}{\log_2(M)}\left(\left(2b\rho_{\rm AR}G_{{\rm ARB}}\right)^{-d} + \left(2b\rho_{\rm AR}G_{{\rm BRA}}\right)^{-d}\right) + o\left(\rho_{\rm AR}^{-d}\right)
\end{equation}
\begin{equation}
d = \min\{M_{\rm A}, M_{\rm B}\} \hspace{2.52 in}
\end{equation}
\begin{equation}
G_{{\rm ARB}} = \left(\frac{a2^{d-1}\eta_{\rm ARB}\Gamma\left(d+\frac{1}{2}\right)}{\sqrt{\pi}d}\right)^{-\frac{1}{d}} \hspace{1.58 in}
\end{equation}
\begin{equation}
G_{{\rm BRA}} = \left(\frac{a2^{d-1}\eta_{\rm BRA}\Gamma\left(d+\frac{1}{2}\right)}{\sqrt{\pi}d}\right)^{-\frac{1}{d}}. \hspace{1.52 in}
\end{equation}\end{footnotesize}\noindent
Note that the diversity order of all five two-way relay systems is \(\min\{M_{\rm A}, M_{\rm B}\}\), and equations (18)-(21) provide tight sum-BER lower-bounds for all five two-way relay protocols.

\section{Multiple Antennas at $\rm R$}

When we consider multiple antennas at \(\rm R\), BF optimization at \(\rm R\) is necessary for the two-slot and first three-slot protocols. In other words, the relay has to simultaneously beamform to both \(\rm A\) and \(\rm B\). In this case, there is no closed-form expression for performance analysis since optimal beamformers cannot be expressed in closed-form. Meanwhile, since BF optimization is not necessary for the second three-slot, first four-slot, and second four-slot protocols, performance analysis with multiple relay antennas can be done with unified received SNRs when \(M_{\rm R} > 1\), which can be represented by equations (1) and (2) with the constants in Table II. Note that this analysis is also applicable for the two-slot and first three-slot protocols as unattainable lower-bounds by assuming that the beamformers are matched, which provides a best case scenario since the beamformers cannot be matched over links \(\rm R\rightarrow A\) and \(\rm R\rightarrow B\) in the same time slot.

In Table II, all values are exact except those denoted by \(D_{\rm ARB,3}\), \(D_{\rm BRA,3}\), \(D_{\rm ARB,4}\), and \(D_{\rm BRA,4}\), which are approximations. To clarify how the approximations in Table II can be obtained, the instantaneous received SNRs are discussed for the second three-slot protocol as an example. The instantaneous received SNRs for the second three-slot protocol at \(\rm A\) and \(\rm B\) are as follows:
\begin{footnotesize}\begin{equation}
\gamma_{\rm BRA} = \gamma_{\rm BRA, 1} + \gamma_{\rm BRA, 2} = \frac{\gamma_{\rm BR}\frac{\gamma_{\rm RA}}{2}}{\gamma_{\rm BR} + \gamma_{\rm AR} + \frac{\gamma_{\rm RA}}{2} + 1} + \frac{\gamma_{\rm BR}\frac{\gamma_{\rm RA}'}{2}}{\gamma_{\rm BR} + \gamma_{\rm AR} + \frac{\gamma_{\rm RA}'}{2} + 1}
\end{equation}
\begin{equation}
\gamma_{\rm ARB} = \gamma_{\rm ARB, 1} + \gamma_{\rm ARB, 2} = \frac{\gamma_{\rm AR}\frac{\gamma_{\rm RB}}{2}}{\gamma_{\rm AR} + \gamma_{\rm BR} + \frac{\gamma_{\rm RB}}{2} + 1} + \frac{\gamma_{\rm AR}\frac{\gamma_{\rm RB}'}{2}}{\gamma_{\rm AR} + \gamma_{\rm BR} + \frac{\gamma_{\rm RB}'}{2} + 1},
\end{equation}\end{footnotesize}\noindent
where \(\gamma_{\rm RA}' = \rho_{\rm RA}\Vert{\mathbf{H}_{\rm RA}\mathbf{f}_{\rm RB}}\Vert^2\) and \(\gamma_{\rm RB}' = \rho_{\rm RB}\Vert{\mathbf{H}_{\rm RB}\mathbf{f}_{\rm RA}}\Vert^2\), which are instantaneous received SNRs with {\it non-matched} BF vectors. Since \(\gamma_{\rm BRA, 1}\) and \(\gamma_{\rm BRA, 2}\) in equation (22) are correlated and \(\gamma_{\rm BRA, 1}\) dominates \(\gamma_{\rm BRA, 2}\), we approximate \(\gamma_{\rm BRA, 2}\) with \(\kappa\gamma_{\rm BRA, 1}\) where \(\kappa := \mathbb{E}[\gamma_{\rm BRA, 2}]/\mathbb{E}[\gamma_{\rm BRA, 1}]\), so that the average values are the same: \(\mathbb{E}[\gamma_{\rm BRA, 2}] = \mathbb{E}[\kappa\gamma_{\rm BRA, 1}]\). Here \(0 < \kappa < 1\) since \(\gamma_{\rm RA}\) is the instantaneous SNR obtained by matched BF, whereas \(\gamma_{\rm RA}'\) results when BF is not matched. This approximation is exact if \(\gamma_{\rm BRA, 2}\) were a constant multiple of \(\gamma_{\rm BRA, 1}\). It becomes tighter as \(M_{\rm R}\) increases because \(\kappa\) becomes smaller as \(M_{\rm R}\) increases, but it is independent of average transmit SNRs, \(M_{\rm A}\), and \(M_{\rm B}\) since they do not have any effect on \(\kappa\), which is checked with numerical investigations. Therefore, \(\gamma_{\rm BRA, 2}\) can be absorbed in \(A_{\rm BRA}\) as in Table II, \(D_{\rm BRA,3} = 1 + \mathbb{E}[\gamma_{\rm BRA, 2}]/\mathbb{E}[\gamma_{\rm BRA, 1}]\). Note that \(D_{\rm BRA,3}\) and \(D_{\rm ARB,3} = 1 + \mathbb{E}[\gamma_{\rm ARB, 2}]/\mathbb{E}[\gamma_{\rm ARB, 1}]\) provide exact performance when \(M_{\rm R} = 1\) such as equations (7) and (18), and they also present a tight performance lower-bound even when \(M_{\rm R} > 1\), which becomes tighter as \(M_{\rm R}\) increases. Similarly, \(D_{\rm ARB,4}\) and \(D_{\rm BRA,4}\) can be obtained for the second four-slot protocol.

\subsection{Performance Analysis}

We now consider performance analysis using the unified received SNRs when \(M_{\rm R} > 1\). Similar to obtaining equation (7) when \(M_{\rm R} = 1\), the distributions of the unified received SNRs for multiple relay antennas should be obtained to calculate sum-BER for \(M_{\rm R} > 1\). The CDFs of \(\Gamma_{\rm BRA}\) and \(\Gamma_{\rm ARB}\) can be obtained as follows (please see Appendix I for derivations):
\begin{footnotesize}\begin{equation}\begin{split}
&F_{\Gamma_{\rm BRA}}(x) = 1 - \sum_{n=1}^{M_{\rm R}}\sum_{m=M_{\rm B}-M_{\rm R}}^{\left(M_{\rm B}+M_{\rm R}\right)n-2n^2}\sum_{k=0}^{m}\sum_{i=1}^{M_{\rm R}}\sum_{j=M_{\rm A}-M_{\rm R}}^{\left(M_{\rm A}+M_{\rm R}\right)i-2i^2}\sum_{p=0}^{k+j}\dbinom{k+j}{p}\frac{2d_{n,m}d_{i,j}}{k!j!\rho_{\rm BR}^{\frac{p+k+1}{2}}\rho_{\rm R}^{\frac{2j+k-p+1}{2}}}\\
&\frac{\left(C_{\rm BRA}n\right)^{\frac{p+k+1}{2}}\left(B_{\rm BRA}i\right)^{\frac{2j+k-p+1}{2}}}{A_{\rm BRA}^{k+j+1}}x^{k+j+1}e^{-\frac{x}{A_{\rm BRA}}\left(\frac{C_{\rm BRA}n}{\rho_{\rm BR}}+\frac{B_{\rm BRA}i}{\rho_{\rm R}}\right)}K_{p-k+1}\left(\frac{2x}{A_{\rm BRA}}\sqrt{\frac{B_{\rm BRA}C_{\rm BRA}ni}{\rho_{\rm BR}\rho_{\rm R}}}\right)
\end{split}\end{equation}
\begin{equation}\begin{split}
&F_{\Gamma_{\rm ARB}}(x) = 1 - \sum_{n=1}^{M_{\rm R}}\sum_{m=M_{\rm A}-M_{\rm R}}^{\left(M_{\rm A}+M_{\rm R}\right)n-2n^2}\sum_{k=0}^{m}\sum_{i=1}^{M_{\rm R}}\sum_{j=M_{\rm B}-M_{\rm R}}^{\left(M_{\rm B}+M_{\rm R}\right)i-2i^2}\sum_{p=0}^{k+j}\dbinom{k+j}{p}\frac{2d_{n,m}d_{i,j}}{k!j!\rho_{\rm AR}^{\frac{p+k+1}{2}}\rho_{\rm R}^{\frac{2j+k-p+1}{2}}}\\
&\frac{\left(C_{\rm ARB}n\right)^{\frac{p+k+1}{2}}\left(B_{\rm ARB}i\right)^{\frac{2j+k-p+1}{2}}}{A_{\rm ARB}^{k+j+1}}
x^{k+j+1}e^{-\frac{x}{A_{\rm ARB}}\left(\frac{C_{\rm ARB}n}{\rho_{\rm AR}}+\frac{B_{\rm ARB}i}{\rho_{\rm R}}\right)}K_{p-k+1}\left(\frac{2x}{A_{\rm ARB}}\sqrt{\frac{B_{\rm ARB}C_{\rm ARB}ni}{\rho_{\rm AR}\rho_{\rm R}}}\right),
\end{split}\end{equation}\end{footnotesize}\noindent
where \(d_{n,m}\) are coefficients given by \cite[eqn.(24)]{dighe03}. Note that equations (24) and (25) are valid when \(M_{\rm A} \ge M_{\rm R}\) and \(M_{\rm B} \ge M_{\rm R}\) even though other cases can be easily handled with minor modifications. For example, \(M_{\rm A}\) and \(M_{\rm R}\) must be switched in equations (24) and (25) when \(M_{\rm A} < M_{\rm R}\). Once equations (24) and (25) are substituted to the second line of equation (4), the sum-BER can be obtained in closed-form similar to equation (7), which are tight sum-BER lower-bounds for the first four-slot, second three-slot, and second four-slot protocols.

\subsection{High-SNR Analysis}

Based on the procedures in Section III.C, we should calculate the \(t_{\rm ARB}\) order derivative of the PDF of \(\lambda_{\rm ARB}\) evaluated at the origin and the \(t_{\rm BRA}\) order derivative of the PDF of \(\lambda_{\rm BRA}\) evaluated at the origin to obtain high-SNR performance when \(M_{\rm R} > 1\). For each path, \(t_{\rm ARB} = t_{\rm BRA} = M_{\rm R}\cdot\min\{M_{\rm A}, M_{\rm B}\} - 1\) since the diversity order of the \(\rm A \rightarrow R \rightarrow B\) and \(\rm B \rightarrow R \rightarrow A\) paths is \(M_{\rm R}\cdot\min\{M_{\rm A}, M_{\rm B}\}\) \cite[eqn.(16)]{kim10_a}. Therefore, the \(t_{\rm ARB}\) and \(t_{\rm BRA}\) order derivatives of the PDFs of \(\lambda_{\rm ARB}\) and \(\lambda_{\rm BRA}\) evaluated at the origin, respectively, can be obtained using the following equations (please see Appendix II for derivation):
\begin{footnotesize}\begin{equation}
f_{\lambda_{\rm AR}}^{(t_{\rm AR})}(0) = \sum_{n=1}^{M_{\rm R}}\sum_{m=M_{\rm A}-M_{\rm R}}^{(M_{\rm A}+M_{\rm R})n-2n^2}d_{n,m}\dbinom{t_{\rm AR}}{m}\left(-1\right)^{t_{\rm AR}+m}\left(\frac{nC_{\rm ARB}}{A_{\rm ARB}}\right)^{t_{\rm AR}+1} \hspace{0.21 in}
\end{equation}
\begin{equation}
f_{\lambda_{\rm RB}}^{(t_{\rm RB})}(0) = \sum_{n=1}^{M_{\rm R}}\sum_{m=M_{\rm B}-M_{\rm R}}^{(M_{\rm B}+M_{\rm R})n-2n^2}d_{n,m}\dbinom{t_{\rm RB}}{m}\left(-1\right)^{t_{\rm RB}+m}\left(\frac{n\rho_{\rm AR}B_{\rm ARB}}{A_{\rm ARB}\rho_{\rm RB}}\right)^{t_{\rm RB}+1}
\end{equation}
\begin{equation}
f_{\lambda_{\rm BR}}^{(t_{\rm BR})}(0) = \sum_{n=1}^{M_{\rm R}}\sum_{m=M_{\rm B}-M_{\rm R}}^{(M_{\rm B}+M_{\rm R})n-2n^2}d_{n,m}\dbinom{t_{\rm BR}}{m}\left(-1\right)^{t_{\rm BR}+m}\left(\frac{n\rho_{\rm AR}C_{\rm BRA}}{A_{\rm BRA}\rho_{\rm BR}}\right)^{t_{\rm BR}+1} \hspace{0.05 in}
\end{equation}
\begin{equation}
f_{\lambda_{\rm RA}}^{(t_{\rm RA})}(0) = \sum_{n=1}^{M_{\rm R}}\sum_{m=M_{\rm A}-M_{\rm R}}^{(M_{\rm A}+M_{\rm R})n-2n^2}d_{n,m}\dbinom{t_{\rm RA}}{m}\left(-1\right)^{t_{\rm RA}+m}\left(\frac{n\rho_{\rm AR}B_{\rm BRA}}{A_{\rm BRA}\rho_{\rm RA}}\right)^{t_{\rm RA}+1},
\end{equation}\end{footnotesize}\noindent
where \(t_{\rm AR} = t_{\rm RA} = M_{\rm A} \cdot M_{\rm R} - 1\) and \(t_{\rm BR} = t_{\rm RB} = M_{\rm B} \cdot M_{\rm R} - 1\) \cite[eqn.(12)]{kim10_a}. Once equations (26)-(29) are substituted into equations (10) and (14), the resulting high-SNR performance using equations (18)-(21) and \(d = M_{\rm R}\cdot\min\{M_{\rm A}, M_{\rm B}\}\) can provide tight sum-BER lower-bounds for the second three-slot, first four-slot, and second four-slot protocols.

\subsubsection{$\alpha$-$\beta$ Optimization}

Following \cite{louie10}, it is possible determine the weighting coefficients used at the relay, \(\alpha\) and \(\beta\), for the first three-slot and second four-slot protocols to minimize instantaneous sum-BERs using brute force search, which is not tractable in closed-form. However, since we are interested in high-SNR performance, we can obtain closed-form expressions using average high-SNR performance in equation (18), especially when \(M_{\rm A} = M_{\rm B} = M_{\rm R} = 1\) as a special case. After every variable is substituted into equation (18) and considering \(\alpha^2 + \beta^2 = 1\), by differentiating equation (18) with respect to \(\beta\), optimal \(\beta\)s for the first three-slot and second four-slot protocols can be obtained, respectively, as follows:
\begin{footnotesize}\begin{equation}
\beta_{\rm three-slot}^2 = \frac{\sqrt{\frac{\rho_{\rm AR}\left(\rho_{\rm AR}+\rho_{\rm RA}\right)}{\rho_{\rm RA}}}}{\sqrt{\frac{\rho_{\rm AR}\left(\rho_{\rm AR}+\rho_{\rm RA}\right)}{\rho_{\rm RA}}} + \sqrt{\frac{\rho_{\rm BR}\left(\rho_{\rm BR}+\rho_{\rm RB}\right)}{\rho_{\rm RB}}}} \hspace{0.15 in}
\end{equation}
\begin{equation}
\beta_{\rm four-slot}^2 = \frac{\sqrt{\frac{\rho_{\rm AR}\left(\rho_{\rm AR}+\rho_{\rm RA}/2\right)}{\rho_{\rm RA}/2}}}{\sqrt{\frac{\rho_{\rm AR}\left(\rho_{\rm AR}+\rho_{\rm RA}/2\right)}{\rho_{\rm RA}/2}} + \sqrt{\frac{\rho_{\rm BR}\left(\rho_{\rm BR}+\rho_{\rm RB}/2\right)}{\rho_{\rm RB}/2}}}
\end{equation}\end{footnotesize}\noindent
Both \(\beta^2\)s become \(\frac{1}{2}\) when \(\rho_{\rm AR} = \rho_{\rm BR} = \rho_{\rm RA} = \rho_{\rm RB}\), while \(\beta^2\)s are bigger than \(\frac{1}{2}\) when \(\rho_{\rm AR} > \rho_{\rm BR}\), which indicates the $\alpha$-$\beta$ optimization is most useful when \(\rho_{\rm AR}\) and \(\rho_{\rm BR}\) are unbalanced. Note that these results are from average high-SNR performance, which leads to worse performance compared with numerically optimizing the instantaneous sum-BERs with respect to \(\beta\). However, equations (30) and (31) do not require instantaneous channel knowledge and can be expressed in closed-form. Note that an implicit equation for optimal \(\beta\) is available even when multiple antennas are considered at all nodes.

\subsubsection{Analytical Gap among Protocols at High-SNR}

We now provide analytical gaps in average SNR for equal \(P_{\rm b}\) between the five protocols at high-SNR. When we compare performance between two protocols, let us denote \(i\) and \(j\) for worse and better protocols in sum-BER, respectively, to make analytical gaps non-negative. Once \(i\) and \(j\) for each protocol are applied to equation (18) and their difference in dB is considered, the analytical gap expression can be obtained as follows:
\begin{footnotesize}\begin{equation}
10\log_{10}\left(\frac{\rho_{\rm AR}^i}{\rho_{\rm AR}^j}\right) = 10\log_{10}\left(\frac{b^j\log_2\left(M^j\right)}{b^i\log_2\left(M^i\right)}\right)
+ \frac{10}{d}\log_{10}\left(\frac{a^i\log_2\left(M^j\right)\left(\eta_{\rm ARB}^i + \eta_{\rm BRA}^i\right)}{a^j\log_2\left(M^i\right)\left(\eta_{\rm ARB}^j + \eta_{\rm BRA}^j\right)}\right).
\end{equation}\end{footnotesize}\noindent
Based on equation (32), we recognize that the analytical gap between protocols \(i\) and \(j\) depends on choice of modulation (i.e. \(a\), \(b\), and \(M\)), diversity order \(d\), and average transmit SNRs and constants from Tables I and II in which need to be substituted to compute \(\eta_{\rm ARB}\) and \(\eta_{\rm BRA}\).

Note that we use QPSK, 8-QAM, and 16-QAM for the two-slot, three-slot, and four-slot protocols, respectively, for rate normalization. Therefore, since \(a\), \(b\) and \(M\) are fixed for all protocols, the analytical gap is mainly determined by the diversity order and the ratio of \(\eta_{\rm ARB}\) and \(\eta_{\rm BRA}\) from equations (10)-(21) as follows:
\begin{footnotesize}\begin{equation}
\frac{\eta_{\rm ARB}^i + \eta_{\rm BRA}^i}{\eta_{\rm ARB}^j + \eta_{\rm BRA}^j} = \frac{\left(\frac{B_{\rm ARB}^i\rho_{\rm AR}^i}{A_{\rm ARB}^i\rho_{\rm RB}^i}\right)^d + \left(\frac{C_{\rm ARB}^i}{A_{\rm ARB}^i}\right)^d + \left(\frac{C_{\rm BRA}^i\rho_{\rm AR}^i}{A_{\rm BRA}^i\rho_{\rm BR}^i}\right)^d + \left(\frac{B_{\rm BRA}^i\rho_{\rm AR}^i}{A_{\rm BRA}^i\rho_{\rm RA}^i}\right)^d}{\left(\frac{B_{\rm ARB}^j\rho_{\rm AR}^j}{A_{\rm ARB}^j\rho_{\rm RB}^j}\right)^d + \left(\frac{C_{\rm ARB}^j}{A_{\rm ARB}^j}\right)^d + \left(\frac{C_{\rm BRA}^j\rho_{\rm AR}^j}{A_{\rm BRA}^j\rho_{\rm BR}^j}\right)^d + \left(\frac{B_{\rm BRA}^j\rho_{\rm AR}^j}{A_{\rm BRA}^j\rho_{\rm RA}^j}\right)^d}.
\end{equation}\end{footnotesize}\noindent
Therefore, the balance between \(\rho_{\rm AR}\) and \(\rho_{\rm BR}\) and the balance between \(M_{\rm A}\) and \(M_{\rm B}\) have an impact on the gap.

For example, when \(\rho_{\rm AR} = \rho_{\rm BR}\) (i.e. balanced), the gap remains the same unless the diversity order is changed. Therefore, if \(M_{\rm A}\) is fixed, the gap increases as \(M_{\rm B}\) increases until \(M_{\rm B}\) reaches to \(M_{\rm A}\), but it remains the same even though \(M_{\rm B}\) increases after \(M_{\rm A} = M_{\rm B}\) due to \(d = M_{\rm R}\cdot\min\{M_{\rm A}, M_{\rm B}\}\). If \(\rho_{\rm AR} \ne \rho_{\rm BR}\) (i.e. unbalanced), \(\alpha\) and \(\beta\) in \(\eta_{\rm ARB}\) and \(\eta_{\rm BRA}\) play important roles on the gap as seen in Section IV.B.1. When the second four-slot protocol with \(\alpha\)-\(\beta\) optimization is compared with other protocols, if \(\rho_{\rm AR} > \rho_{\rm BR}\) with \(M_{\rm A} = M_{\rm B}\), the gap increases as \(\rho_{\rm AR}\) increases due to the benefit of \(\alpha\)-\(\beta\) optimization. However, since \(\beta^2 \approx 0\) when \(M_{\rm A} < M_{\rm B}\) regardless of \(\rho_{\rm AR}\) and \(\rho_{\rm BR}\), the combination of \(\rho_{\rm AR} > \rho_{\rm BR}\) and \(M_{\rm A} < M_{\rm B}\) removes an advantage of the \(\alpha\)-\(\beta\) optimization, so that other protocols have better performance than the second four-slot protocol in this case. Therefore, the \(\alpha\)-\(\beta\) optimization can be useful when \(\rho_{\rm AR} \ne \rho_{\rm BR}\) with careful consideration of \(M_{\rm A}\) and \(M_{\rm B}\).

\section{Numerical and Simulation Results}

In Monte-Carlo simulations, the transmitted symbol is QPSK, 8-QAM, or 16-QAM modulated for two-slot, three-slot, four-slot protocols, respectively, for rate normalization. Zero mean and unit variance are used to model the Rayleigh block fading channel. The distance between \(\rm A\) and \(\rm R\) is set as a reference \(d_0\) whereas the distance between \(\rm A\) and \(\rm B\) is \(d\). Therefore, once \(d_0\) is determined, \(10\log_{10}(\rho_{\rm BR}) = 10\log_{10}(\rho_{\rm AR}) - 10\gamma\log_{10}((1-d_0)/d_0)\), where \(\gamma\) is the path-loss exponent of the simplified path-loss model in \cite{goldsmith05}. Note that average transmit power per node is normalized in unified received SNR expressions for fair comparison among all protocols.

\subsection{Accuracy of Analysis}

This subsection illustrates the accuracy of our analysis in equations (7) and (18) with \(M_{\rm R} = 1\), and the analysis using equations (24), (25), and (26)-(29) with \(M_{\rm R} > 1\). Figures 3 and 4 show \(2\times1\times2\) and \(2\times2\times2\) AF MIMO BF two-way relay network sum-BER performance when both average transmit SNRs are balanced (i.e. \(\rho_{\rm AR} = \rho_{\rm BR}\) due to \(d_0 = 0.5\)), respectively. All simulation curves in Figures 3 and 4 are from Monte-Carlo simulations. All analytical curves of five protocols are from equation (7) and using equations (24) and (25) with proper constants given in Tables I and II. All high-SNR analytical curves are from equation (18) and using equations (26)-(29) with related constants in Tables I and II. Our analysis including high-SNR analysis matches exactly with Monte-Carlo simulations at high-SNR in Figures 3 and 4. Note that sum-BER performance in equation (7) and using equations (24) and (25) provides tight lower-bounds to equation (3).

\subsection{$\alpha$-$\beta$ Optimization}

This subsection shows $\alpha$-$\beta$ optimization related figures. Figure 5 shows the optimal $\beta^2$ for the first three-slot and second four-slot protocols average sum-BER at high-SNR using equation (18) for \(1\times1\times1\) AF two-way relay network performance with \(\rho_{\rm AR} = \rho_{\rm RA} = \rho_{\rm RB} = 40\) dB when average transmit SNRs are unbalanced (i.e. \(\rho_{\rm AR} \ne \rho_{\rm BR}\) due to \(d_0 \ne 0.5\)) to show the accuracy of equations (30) and (31). Using the same setup, analytical results in equations (30) and (31) present \(\beta^2 = 0.82915\) and \(\beta^2 = 0.85159\) for the optimal \(\beta^2\) in the first three-slot and second four-slot protocols, respectively.

Figure 6 shows \(2\times1\times2\) AF MIMO BF two-way relay network performance in sum-BER when average transmit SNRs are unbalanced. All simulation curves in Figure 6 are from numerical simulations using equation (3), where the optimal $\beta$s are selected based on the instantaneous error rate expression in equation (3) and depends on the channel realizations. All analytical curves of two protocols are from equation (7) with proper constants. All high-SNR analytical curves are from equation (18) with related constants. $\beta^2 = 0.87196$ and $\beta^2 = 0.88471$ are used for optimal values at high-SNR using equation (18) for the first three-slot and second four-slot protocols, respectively. The optimal $\beta$s are chosen to minimize average high-SNR in this case. About 1 dB performance gap exists between the case of selection of $\beta$ based on instantaneous channel realizations versus selection of $\beta$ based on high SNR sum-BER expressions similar to equations (30) and (31).

\subsection{Comparisons of Protocols}

This subsection compares sum-BER performance among five relaying protocols. Note that \(\alpha\)-\(\beta\) optimization is performed when average transmit SNRs are unbalanced, and BF optimization, using the gradient algorithm in \cite{khoshnevis08} for the first 3-slot protocol and the iterative minimum sum-MSE (MSMSE) from \cite{lee08, lee10} for the 2-slot protocol, is conducted when multiple relay antennas are used. Figure 7 shows \(2\times2\times2\) AF MIMO BF two-way relay network performance comparison among five protocols when average transmit SNRs are balanced. All simulation curves are from numerical simulations with \(\Gamma_{\rm ARB}\) and \(\Gamma_{\rm BRA}\) for fair comparison, and all analytical curves are using equations (24) and (25) with proper constants. Note that the two-slot and first three-slot protocols need to find optimal beamformers for minimum sum-BER. Our proposed three-slot protocol with normalized rate outperforms all other protocols at high-SNR in Figure 7.

Figure 8 shows \(2\times1\times2\) AF MIMO BF two-way relay network performance comparison when \(d_0 = 0.3\). All simulation curves are from numerical simulations using equation (3) with \(\alpha\)-\(\beta\) optimization. All analytical curves are from equation (7) with proper constants. The first three-slot and second four-slot protocols find optimal \(\alpha\) and \(\beta\) using the instantaneous approach. Our proposed four-slot protocol with optimal \(\alpha\) and \(\beta\) and normalized rate outperforms all other protocols at high-SNR in Figure 8.

The analytical high-SNR gaps between five protocols for three scenarios based on equations (18) and (26)-(29) are given in Table III. All gaps are from the best protocol for each scenario in dB. For example, the best protocol in sum-BER for \(2\times1\times2\) AF MIMO BF two-way relay networks when transmit SNRs are balanced is the two-slot protocol, and the gap from the two-slot protocol to the second three-slot protocol is 0.6608 dB. Note that the proposed four-slot protocol is the best protocol for \(2\times1\times2\) AF MIMO BF two-way relay networks when transmit SNRs are unbalanced, and the proposed three-slot protocol is the best protocol for \(2\times2\times2\) AF MIMO BF two-way relay networks when transmit SNRs are balanced.

\section{Conclusions}

Unified performance analysis has been conducted for AF MIMO BF two-way relay networks with five different relaying protocols using two, three, or four time slots. We first have introduced novel ``second three-slot" and ``second four-slot" protocols suitable for BF and better sum-BER performance. Novel closed-form unified sum-BER expressions have been presented with corresponding closed-form unified CDFs. Furthermore, new closed-form unified high-SNR performance expressions have been provided for simplicity and mathematical tractability, and the analytical high-SNR gap expression is provided.

Based on analytical and simulation results, we have investigated the performance of five different protocols with two, three, or four time slots using the sum-BER metric. As a result, we can conclude that the proposed three-slot protocol outperforms all other protocols at high-SNR when multiple relay antennas are used, and the proposed four-slot protocol outperforms all other protocols at high-SNR when average transmit SNRs are unbalanced. Therefore, we can say that the proposed protocols are a good alternative to the two-slot protocol when multiple relay antennas are used and average transmit SNRs are unbalanced.

\section*{Appendix I: Derivations of Equations (5), (6), (24), and (25)}

This appendix derives the CDFs of \(\Gamma_{\rm ARB}\) and \(\Gamma_{\rm BRA}\) with a general \(M_{\rm R}\) so that it covers equations (5), (6), (24), and (25). We derive the CDF of \(\Gamma_{\rm ARB}\) first and discuss the CDF of \(\Gamma_{\rm BRA}\) later. For the CDF of \(\Gamma_{\rm ARB}\), the following procedures can be used by the definitions of CDF and complementary CDF (CCDF):
\begin{footnotesize}\begin{equation}\begin{split}
&F_{\Gamma_{\rm ARB}}(x) = \int_0^{\infty}Pr\left(\frac{A_{\rm ARB}\gamma_{\rm AR}y}{B_{\rm ARB}\gamma_{\rm AR}+C_{\rm ARB}y} \le x\right)f_{\gamma_{\rm RB}}(y)dy\\
&\hspace{0.55 in}= 1 - \int_0^{\infty}\bar{F}_{\gamma_{\rm AR}}\left(\frac{C_{\rm ARB}x\left(w+B_{\rm ARB}x\right)}{A_{\rm ARB}w}\right)f_{\gamma_{\rm RB}}\left(\frac{w+B_{\rm ARB}x}{A_{\rm ARB}}\right)\frac{dw}{A_{\rm ARB}},
\end{split}\end{equation}\end{footnotesize}\noindent
where \(\bar{F}_{\gamma_{\rm AR}}(x)\) is the CCDF of \(\gamma_{\rm AR}\), which \(\bar{F}_{\gamma_{\rm AR}}(x) = 1 - F_{\gamma_{\rm AR}}(x)\). Since the CDF of \(\gamma_{\rm AR}\) and the PDF of \(\gamma_{\rm RB}\) are given by \cite[eqns.(24)-(25)]{kim10_a}
\begin{footnotesize}\begin{equation}
F_{\gamma_{\rm AR}}(x) = 1 - \sum_{n=1}^{M_{\rm R}}\sum_{m=M_{\rm A}-M_{\rm R}}^{\left(M_{\rm A}+M_{\rm R}\right)n-2n^2}\sum_{k=0}^{m}\frac{d_{n,m}(nx)^ke^{-nx/\rho_{\rm AR}}}{k!\rho_{\rm AR}^k}, \hspace{0.15 in} x>0
\end{equation}
\begin{equation}
f_{\gamma_{\rm RB}}(x) = \sum_{i=1}^{M_{\rm R}}\sum_{j=M_{\rm B}-M_{\rm R}}^{\left(M_{\rm B}+M_{\rm R}\right)i-2i^2}\frac{d_{i,j}i^{j+1}x^je^{-ix/\rho_{\rm R}}}{j!\rho_{\rm R}^{j+1}}, \hspace{0.15 in} x>0. \hspace{0.65 in}
\end{equation}\end{footnotesize}\noindent
Equation (25) can be acquired after complicated mathematical manipulations if equations (35) and (36) are substituted to the last line of equation (34). Using similar procedures, equation (24) can also be obtained using the corresponding constants and subscripts. Once \(M_{\rm R} = 1\) is applied to equations (24) and (25), equations (5) and (6) can be attained.

\section*{Appendix II: Derivations of Equations (10), (14), and (26)-(29)}

This appendix derives the \(t_{\rm ARB}\) and \(t_{\rm BRA}\) order derivatives of the PDFs of \(\lambda_{\rm ARB}\) and \(\lambda_{\rm BRA}\) evaluated at the origin, respectively, with a general \(M_{\rm R}\) so that it covers all cases. We derive the \(t_{\rm ARB}\) order derivative of the PDF of \(\lambda_{\rm ARB}\) and evaluate it at the origin first, and then we discuss the \(t_{\rm BRA}\) order derivative of the PDF of \(\lambda_{\rm BRA}\) evaluated at the origin later. To acquire the \(t_{\rm ARB}\) order derivative of the PDF of \(\lambda_{\rm ARB}\), we need to obtain the PDF of \(\lambda_{\rm ARB}\). Since \(\lambda_{\rm ARB} = \Gamma_{\rm ARB}/\rho_{\rm AR}\), we can easily find the PDF of \(\lambda_{\rm ARB}\) if the PDF of \(\Gamma_{\rm ARB}\) is given. From equation (2), \(\Gamma_{\rm ARB}\) can be rewritten as
\begin{footnotesize}\begin{equation}
\Gamma_{\rm ARB} = \frac{A_{\rm ARB}\gamma_{\rm AR}\gamma_{\rm RB}}{B_{\rm ARB}\gamma_{\rm AR}+ C_{\rm ARB}\gamma_{\rm RB}} = \frac{A_{\rm ARB}}{B_{\rm ARB}C_{\rm ARB}}\frac{B_{\rm ARB}C_{\rm ARB}\gamma_{\rm AR}\gamma_{\rm RB}}{B_{\rm ARB}\gamma_{\rm AR} + C_{\rm ARB}\gamma_{\rm RB}} = \frac{A_{\rm ARB}}{B_{\rm ARB}C_{\rm ARB}}W,
\end{equation}\end{footnotesize}\noindent
where \(W := B_{\rm ARB}C_{\rm ARB}\gamma_{\rm AR}\gamma_{\rm RB}/\left(B_{\rm ARB}\gamma_{\rm AR} + C_{\rm ARB}\gamma_{\rm RB}\right)\), which is the received SNR of a two-hop relay system when the noise variance of the first hop is removed.

Since we consider high-SNR, \(W\) can be approximated by \(\min\{B_{\rm ARB}\gamma_{\rm AR}, C_{\rm ARB}\gamma_{\rm RB}\}\) \cite{louie10}. Based on the identity for the minimum of two independent RVs in \cite[eqn.(6.58)]{papoulis02}, the PDF of \(W\) can be approximated at high-SNR as
\begin{footnotesize}\begin{equation}\begin{split}
&f_W(x) \approx f_{B_{\rm ARB}\gamma_{\rm AR}}(x)\bar{F}_{C_{\rm ARB}\gamma_{\rm RB}}(x) + f_{C_{\rm ARB}\gamma_{\rm RB}}(x)\bar{F}_{B_{\rm ARB}\gamma_{\rm AR}}(x) \\
&\hspace{0.35 in} = \sum_{n=1}^{M_{\rm R}}\sum_{m=M_{\rm A}-M_{\rm R}}^{\left(M_{\rm A}+M_{\rm R}\right)n-2n^2}\sum_{i=1}^{M_{\rm R}}\sum_{j=M_{\rm B}-M_{\rm R}}^{\left(M_{\rm B}+M_{\rm R}\right)i-2i^2}\sum_{p=0}^{j}\frac{d_{n,m}d_{i,j}n^{m+1}i^px^{m+p}e^{-x\left(\frac{n}{B_{\rm ARB}\rho_{\rm AR}}+\frac{i}{C_{\rm ARB}\rho_{\rm R}}\right)}}{m!p!\left(B_{\rm ARB}\rho_{\rm AR}\right)^{m+1}\left(C_{\rm ARB}\rho_{\rm R}\right)^p} \\
&\hspace{0.35 in} + \sum_{i=1}^{M_{\rm R}}\sum_{j=M_{\rm B}-M_{\rm R}}^{\left(M_{\rm B}+M_{\rm R}\right)i-2i^2}\sum_{n=1}^{M_{\rm R}}\sum_{m=M_{\rm A}-M_{\rm R}}^{\left(M_{\rm A}+M_{\rm R}\right)n-2n^2}\sum_{k=0}^{m}\frac{d_{n,m}d_{i,j}n^ki^{j+1}x^{k+j}e^{-x\left(\frac{n}{B_{\rm ARB}\rho_{\rm AR}}+\frac{i}{C_{\rm ARB}\rho_{\rm R}}\right)}}{k!j!\left(B_{\rm ARB}\rho_{\rm AR}\right)^k\left(C_{\rm ARB}\rho_{\rm R}\right)^{j+1}}.
\end{split}\end{equation}\end{footnotesize}\noindent

Using the identity of \cite[eqn.(6.5)]{papoulis02}, the PDF of \(\lambda_{\rm ARB}\) can be approximated at high-SNR as \(f_{\lambda_{\rm ARB}}(x) \approx B_{\rm ARB}C_{\rm ARB}\rho_{\rm AR}f_{W}\left(B_{\rm ARB}C_{\rm ARB}\rho_{\rm AR}x/A_{\rm ARB}\right)/A_{\rm ARB}\) since \(\lambda_{\rm ARB} = \Gamma_{\rm ARB}/\rho_{\rm AR}\). Once \(f_{\lambda_{\rm ARB}}(x)\) is differentiated \(t_{\rm ARB}\) times and evaluated at the origin for each case (i.e. \(M_{\rm A} > M_{\rm B}\), \(M_{\rm A} < M_{\rm B}\), and \(M_{\rm A} = M_{\rm B}\)), equation (10) can be obtained, where \(f_{\lambda_{\rm AR}}^{(t_{\rm AR})}(0)\) and \(f_{\lambda_{\rm RB}}^{(t_{\rm RB})}(0)\) are given in equations (26) and (27), respectively. Once \(M_{\rm R} = 1\) is applied, equations (11) and (12) can be attained. Using similar procedures, equations (14)-(16) can be obtained.

\bibliographystyle{IEEEtran}
\bibliography{Hyunjun_References}

\begin{table}[htbp]
\caption{The coefficients for equations (1) and (2) when $M_{\rm R} = 1$}
\centering
\begin{tabular}{|c||c|c|c||c|c|c|} \hline
Constants & $A_{\rm BRA}$ & $B_{\rm BRA}$ & $C_{\rm BRA}$ & $A_{\rm ARB}$ & $B_{\rm ARB}$ & $C_{\rm ARB}$ \\ \hline
2-slot & 1 & 1 & $1+\frac{\rho_{\rm AR}}{\rho_{\rm RA}}$ & 1 & 1 & $1+\frac{\rho_{\rm BR}}{\rho_{\rm RB}}$ \\ \hline
First 3-slot & $\beta^2$ & $\beta^2$ & $1+\frac{\alpha^2\rho_{\rm AR}}{\rho_{\rm RA}}$ & $\alpha^2$ & $\alpha^2$ & $1+\frac{\beta^2\rho_{\rm BR}}{\rho_{\rm RB}}$ \\ \hline
First 4-slot & $\frac{1}{2}$ & 1 & $\frac{1}{2}$ & $\frac{1}{2}$ & 1 & $\frac{1}{2}$ \\ \hline
Second 3-slot & 1 & 1 & $\frac{1}{2}+\frac{\rho_{\rm AR}}{\rho_{\rm RA}}$ & 1 & 1 & $\frac{1}{2}+\frac{\rho_{\rm BR}}{\rho_{\rm RB}}$ \\ \hline
Second 4-slot & $\beta^2$ & $\beta^2$ & $\frac{1}{2}+\frac{\alpha^2\rho_{\rm AR}}{\rho_{\rm RA}}$ & $\alpha^2$ & $\alpha^2$ & $\frac{1}{2}+\frac{\beta^2\rho_{\rm BR}}{\rho_{\rm RB}}$ \\ \hline
\end{tabular}
\end{table}\noindent

\begin{table}[htbp]
\caption{The coefficients for equations (1) and (2) when $M_{\rm R} > 1$}
\centering
\begin{tabular}{|c||c|c|c||c|c|c|} \hline
Constants & $A_{\rm BRA}$ & $B_{\rm BRA}$ & $C_{\rm BRA}$ & $A_{\rm ARB}$ & $B_{\rm ARB}$ & $C_{\rm ARB}$ \\ \hline
2-slot & 1 & 1 & $1+\frac{\rho_{\rm AR}}{\rho_{\rm RA}}$ & 1 & 1 & $1+\frac{\rho_{\rm BR}}{\rho_{\rm RB}}$ \\ \hline
First 3-slot & $\beta^2$ & $\beta^2$ & $1+\frac{\alpha^2\rho_{\rm AR}}{\rho_{\rm RA}}$ & $\alpha^2$ & $\alpha^2$ & $1+\frac{\beta^2\rho_{\rm BR}}{\rho_{\rm RB}}$ \\ \hline
First 4-slot & $\frac{1}{2}$ & 1 & $\frac{1}{2}$ & $\frac{1}{2}$ & 1 & $\frac{1}{2}$ \\ \hline
Second 3-slot & $\frac{D_{\rm BRA,3}}{2}$ & 1 & $\frac{1}{2}+\frac{\rho_{\rm AR}}{\rho_{\rm RA}}$ & $\frac{D_{\rm ARB,3}}{2}$ & 1 & $\frac{1}{2}+\frac{\rho_{\rm BR}}{\rho_{\rm RB}}$ \\ \hline
Second 4-slot & $\beta^2\frac{D_{\rm BRA,4}}{2}$ & $\beta^2$ & $\frac{1}{2}+\frac{\alpha^2\rho_{\rm AR}}{\rho_{\rm RA}}$ & $\alpha^2\frac{D_{\rm ARB,4}}{2}$ & $\alpha^2$ & $\frac{1}{2}+\frac{\beta^2\rho_{\rm BR}}{\rho_{\rm RB}}$ \\ \hline
\end{tabular}
\end{table}\noindent

\begin{table}[htbp]
\caption{The analytical high-SNR gaps in equation (32) between five protocols in dB}
\centering
\begin{tabular}{|c|c||c|c||c|c|} \hline
\multicolumn{4}{|c||}{$2\times1\times2$ Two-way Relay} & \multicolumn{2}{|c|}{$2\times2\times2$ Two-way Relay} \\ \hline
\multicolumn{2}{|c||}{Balanced SNR} & \multicolumn{2}{|c||}{Unbalanced SNR} & \multicolumn{2}{|c|}{Balanced SNR} \\ \hline
Best Protocol & 2-slot & Best Protocol & Second 4-slot & Best Protocol & Second 3-slot \\ \hline
Gap to First 3-slot & 3.1014 & Gap to 2-slot & 1.7106 & Gap to 2-slot & 0.8412 \\ \hline
Gap to Second 3-slot & 0.6608 & Gap to First 3-slot & 0.9651 & Gap to First 3-slot & 2.9071 \\ \hline
Gap to First 4-slot & 3.3547 & Gap to Second 3-slot & 2.0607 & Gap to First 4-slot & 2.1083 \\ \hline
Gap to Second 4-slot & 3.3547 & Gap to First 4-slot & 1.1622 & Gap to Second 4-slot & 2.9495 \\ \hline
\end{tabular}
\end{table}\noindent
\begin{figure}[htbp]
\begin{center}
  \includegraphics[width=0.65\textwidth,height=0.3\textwidth]{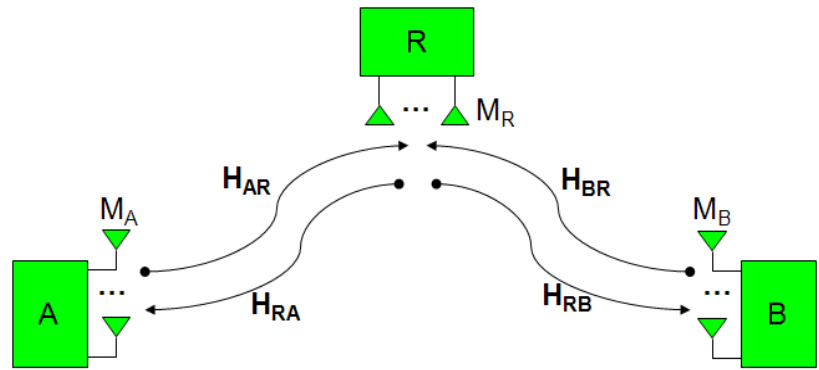}
  \caption{System Model of Two-Hop MIMO Two-Way Relay Networks.}
  \label{fig:System Model}
\end{center}
\end{figure}

\begin{figure}[htbp]
\begin{center}
  \includegraphics[width=0.9\textwidth,height=0.7\textwidth]{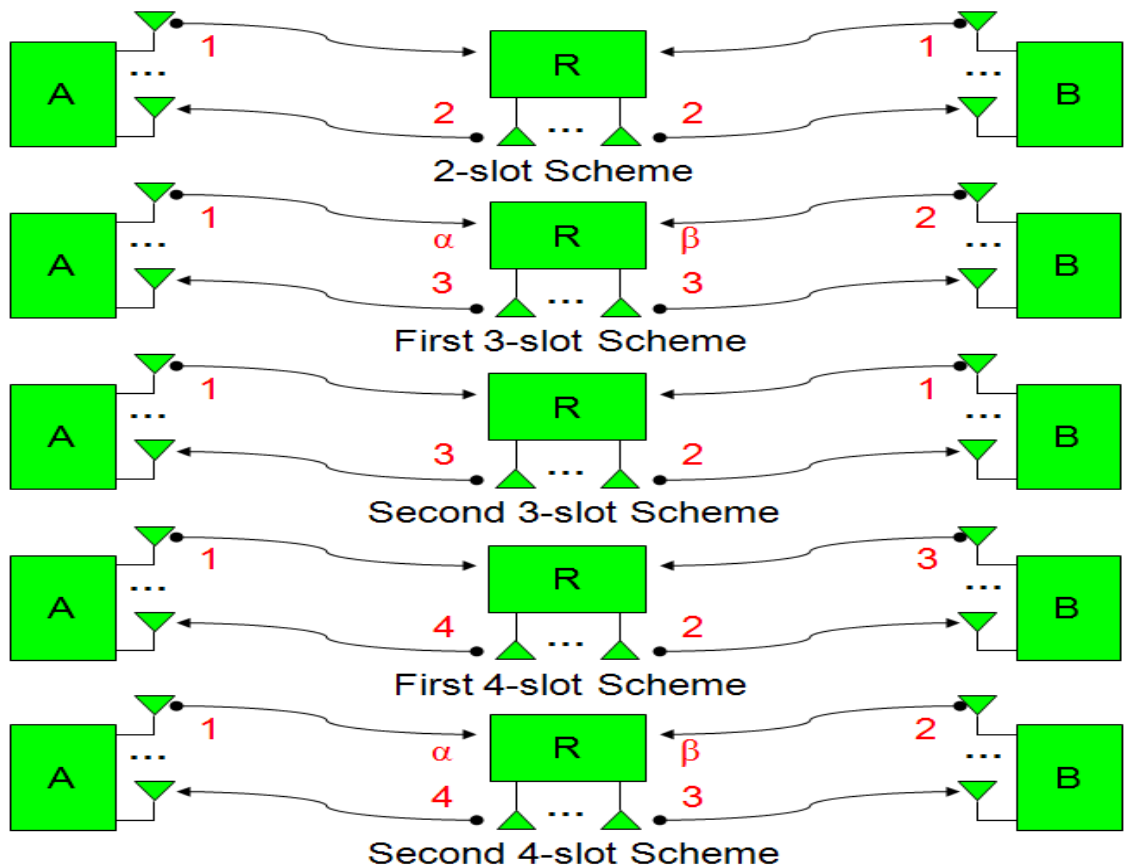}
  \caption{Transmission Protocols for Two-Way Relay Networks.}
  \label{fig:System Model}
\end{center}
\end{figure}

\begin{figure}[htbp]
\begin{center}
  \includegraphics[width=0.8\textwidth,height=0.55\textwidth]{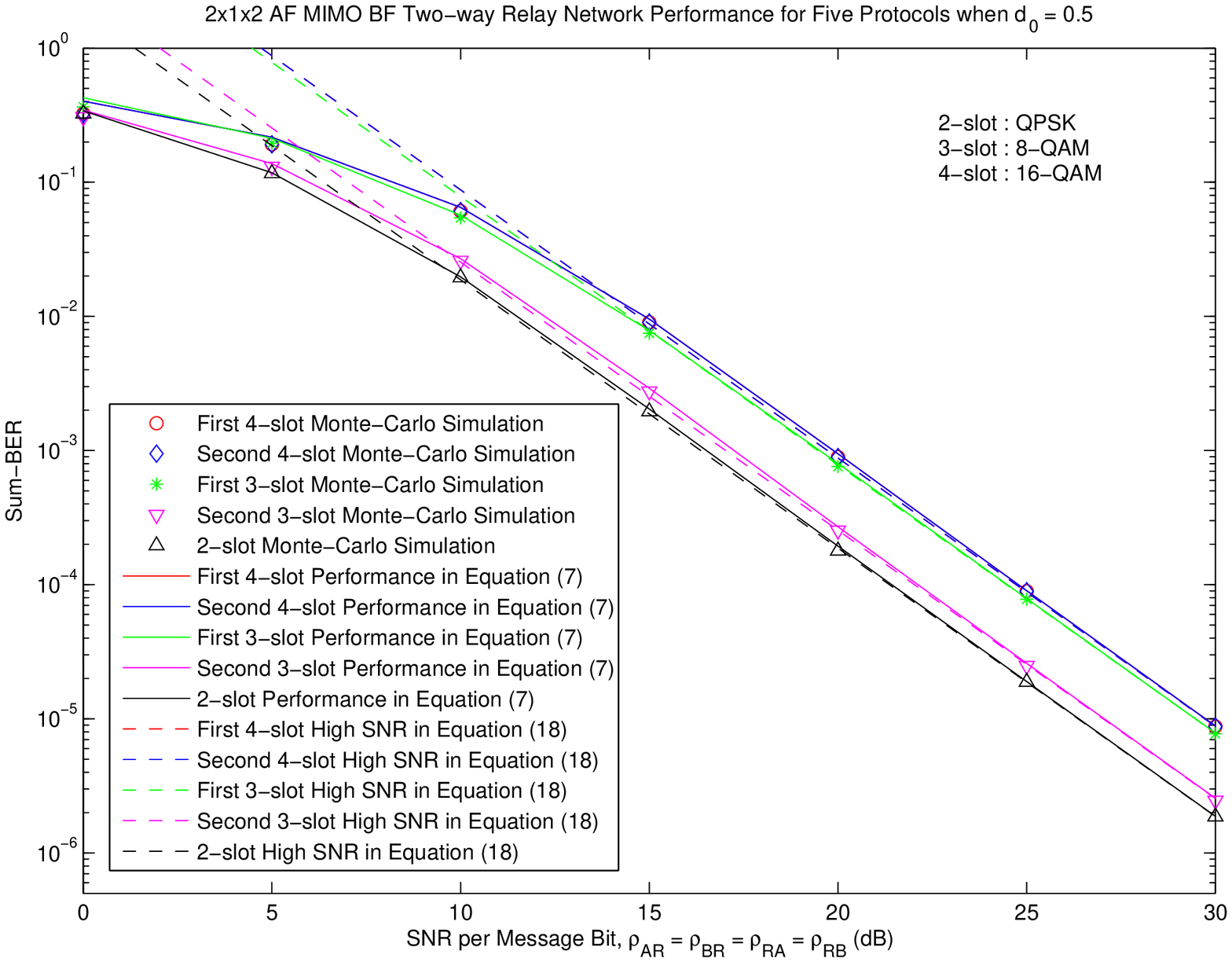}
  \caption{\(2\times1\times2\) AF MIMO BF Two-Way Relay Network Performance in Sum-BER when \(d_0 = 0.5\).}
\end{center}
\end{figure}

\begin{figure}[htbp]
\begin{center}
  \includegraphics[width=0.8\textwidth,height=0.55\textwidth]{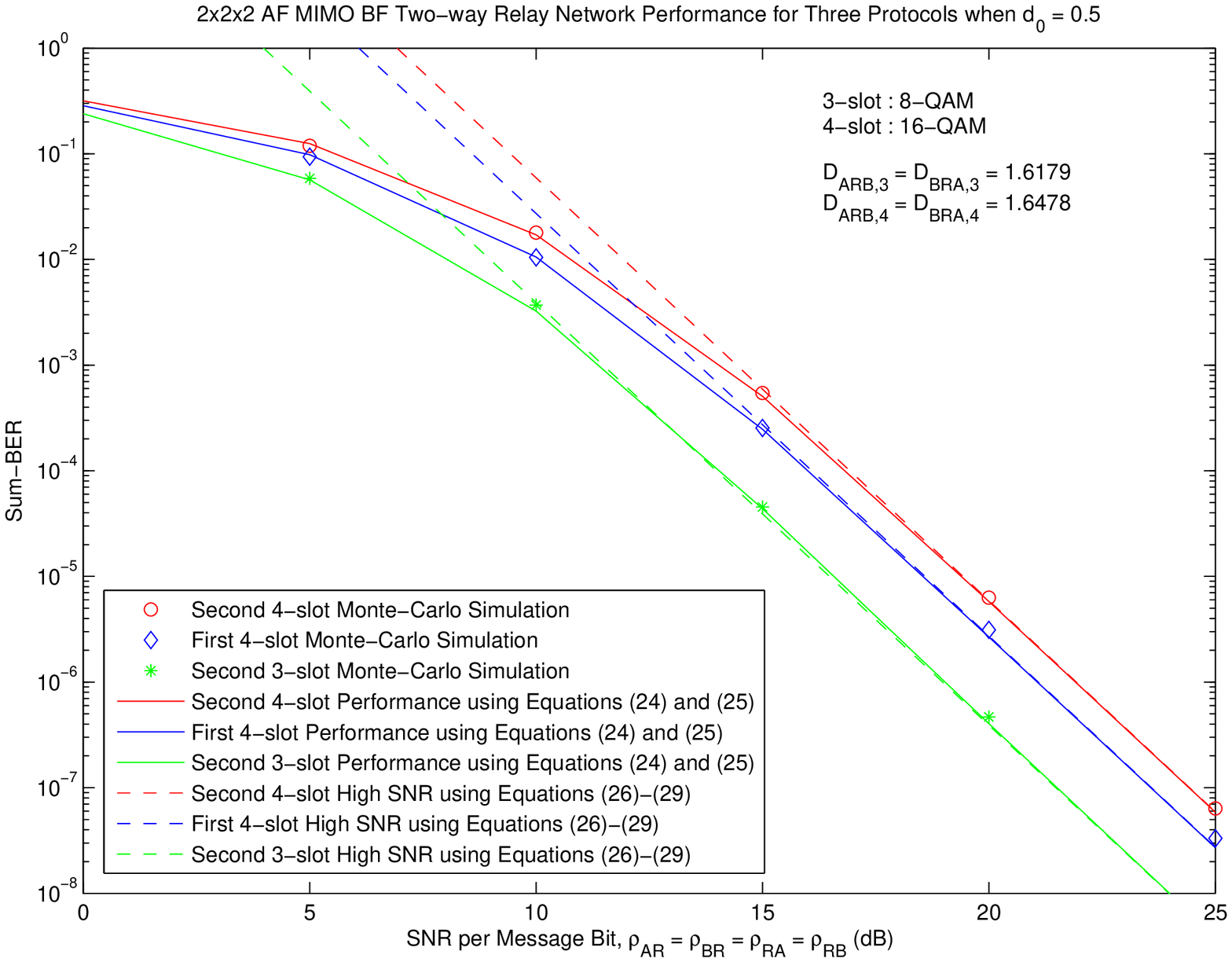}
  \caption{\(2\times2\times2\) AF MIMO BF Two-Way Relay Network Performance in Sum-BER when \(d_0 = 0.5\).}
\end{center}
\end{figure}

\begin{figure}[htbp]
\begin{center}
  \includegraphics[width=0.8\textwidth,height=0.55\textwidth]{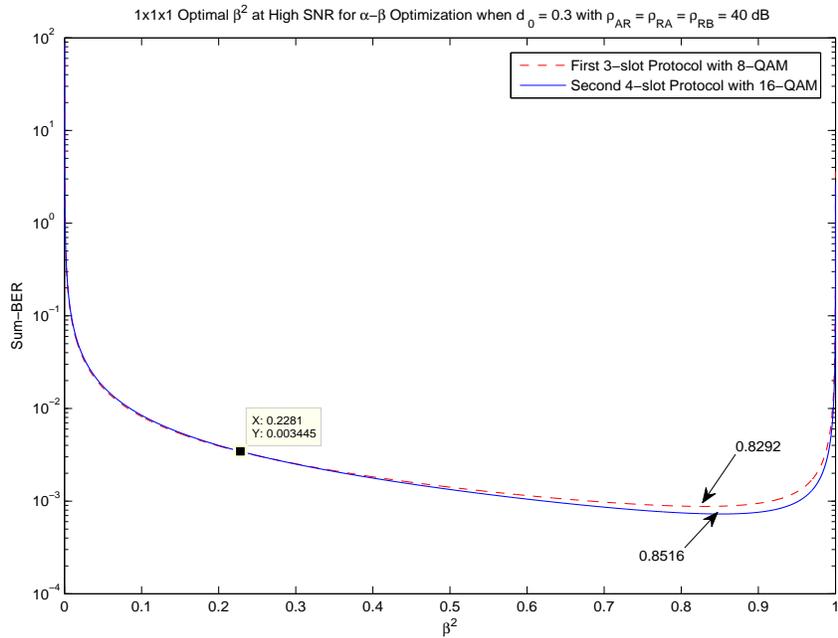}
  \caption{Optimal $\beta^2$ at High-SNR for \(1\times1\times1\) AF Two-Way Relay Network Performance with \(\rho_{\rm AR} = 40\) dB when \(d_0 = 0.3\).}
\end{center}
\end{figure}

\begin{figure}[htbp]
\begin{center}
  \includegraphics[width=0.8\textwidth,height=0.55\textwidth]{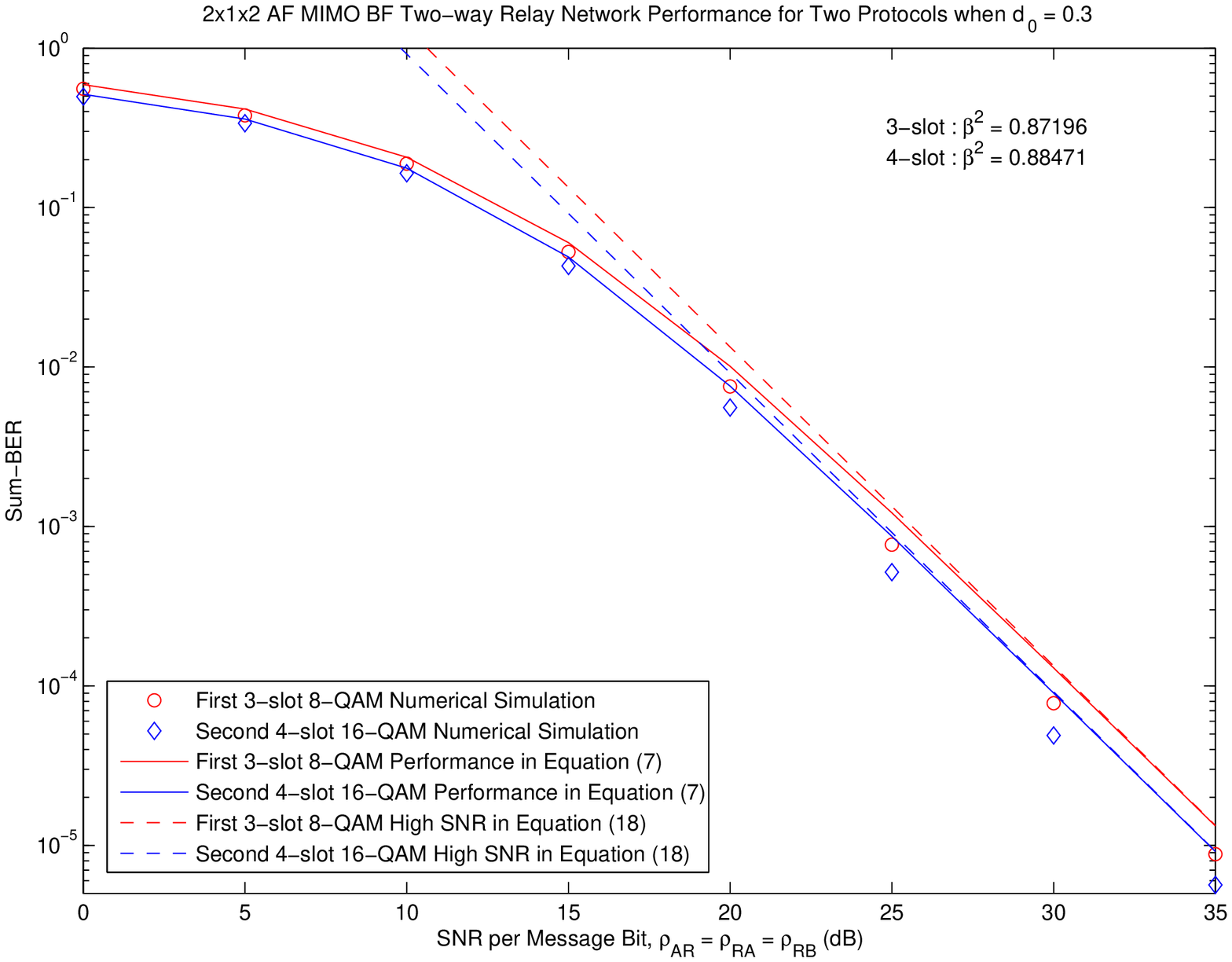}
  \caption{\(2\times1\times2\) AF MIMO BF Two-Way Relay Network Performance in Sum-BER when \(d_0 = 0.3\).}
\end{center}
\end{figure}

\begin{figure}[htbp]
\begin{center}
  \includegraphics[width=0.8\textwidth,height=0.55\textwidth]{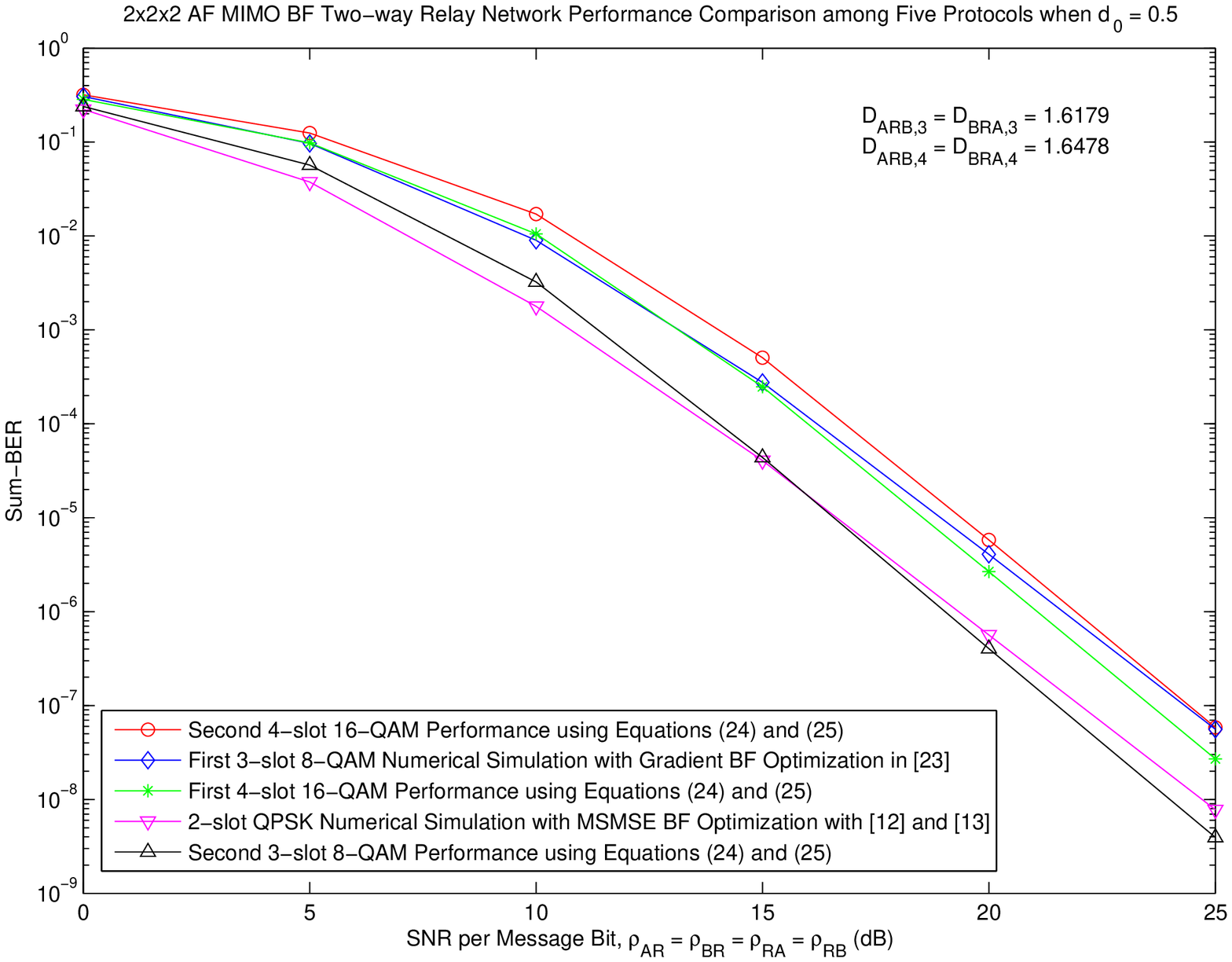}
  \caption{\(2\times2\times2\) AF MIMO BF Two-Way Relay Network Performance in Sum-BER when \(d_0 = 0.5\).}
\end{center}
\end{figure}

\begin{figure}[htbp]
\begin{center}
  \includegraphics[width=0.8\textwidth,height=0.55\textwidth]{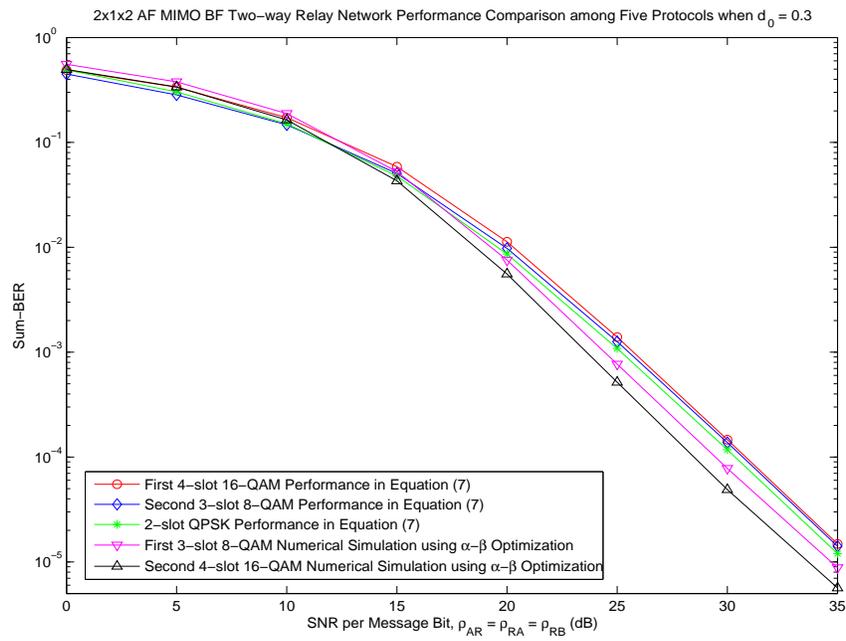}
  \caption{\(2\times1\times2\) AF MIMO BF Two-Way Relay Network Performance Comparison when \(d_0 = 0.3\).}
\end{center}
\end{figure}

\end{document}